\documentclass[prb,aps,12pt,a4paper,showpacs]{revtex4}

\usepackage{epsfig}

\begin{document}

\title{\bf Virial expansion for almost diagonal random matrices}

\author{Oleg Yevtushenko}
\email{bom@ictp.trieste.it}
\affiliation{The Abdus Salam ICTP, Strada Costiera 11, 34100, Trieste, Italy}

\author{Vladimir Kravtsov}
\email{kravtsov@ictp.trieste.it}
\affiliation{ The Abdus Salam ICTP, Strada Costiera 11, 34100, Trieste, Italy \\
              Landau Institute for Theoretical Physics, 2 Kosygina st.,
              117940 Moscow, Russia
            }

\date{\today}

\begin{abstract}
Energy level statistics of Hermitian random matrices $ \, \hat H
\,$ with Gaussian independent random entries $H_{i\geq j}$ is
studied for a generic ensemble of almost diagonal random matrices
with $ \, \langle|H_{ii}|^{2} \rangle \sim 1 \, $ and  $ \, \langle|H_{i\neq
j}|^{2} \rangle= b\, {\cal F}(|i-j|) \ll 1 $. We perform a regular
expansion of the spectral form-factor $ \, K(\tau) = 1 + b\,K_{1}(\tau) + 
b^{2}\,K_{2}(\tau) + \ldots \, $ in powers of $ \, b \ll 1 \, $ 
with the coefficients $ \, K_{m}(\tau) \, $ that take into account
interaction of $ \, (m+1) \, $ energy levels. To calculate 
$ \, K_{m}(\tau) $, we develop a diagrammatic technique
which is based on the Trotter formula and on the combinatorial problem of 
graph edges coloring with $ \, (m+1) \, $ colors. Expressions for $ \, K_{1}(\tau) \, $ 
and $ \, K_{2}(\tau) \, $ in terms of infinite series are found for a generic 
function $ \, {\cal F}(|i-j|) \, $ in the Gaussian Orthogonal Ensemble (GOE), 
the Gaussian Unitary Ensemble (GUE) and in the crossover between them (the almost 
unitary Gaussian ensemble). The Rosenzweig-Porter and power-law banded matrix ensembles 
are considered as examples.
\end{abstract}

{}

\maketitle

\section{Introduction}

Random Matrix Theory (RMT) has proven to be a universal
formalism to describe a great variety of complex systems ranging from
nuclei to mesoscopic quantum dots \cite{GlAl} to chaotic systems
\cite{KickRot}. Of particularly wide application is the Wigner-Dyson RMT
\cite{WigDy,Mehta,rev}. It is the statistical theory of eigenvalues and
eigenfunctions of a random Hermitian matrix $ \, \hat H \, $ whose
entries $H_{i\geq j}$  fluctuate as independent Gaussian random variables
with zero mean $\langle H_{ij} \rangle=0$ and a constant variance $\langle
|H_{ij}|^{2} \rangle={\rm const}$.
There are three Dyson symmetry classes labeled by $\beta$: orthogonal ($\beta=1$),
unitary ($\beta=2$) and symplectic ($\beta=4$) that correspond to real, generic
complex, and real quaternionic Hermitian matrices. This theory is extremely successful
in describing the spectrum of complex nuclei and the statistics of various observable
quantities in mesoscopic quantum dots \cite{GlAl,AlVav,AlVavAmb}.

The key property of the Wigner-Dyson RMT is that the variance $\langle |H_{ij}|^{2}
\rangle$ does not change with increasing the distance $|i-j|$ from the diagonal.
In order to illustrate the properties of eigenvectors one can invoke one-dimensional
chain of sites $1<i<N$ ($N$ is the matrix size)
with on-site energies $H_{ii}$ and hopping matrix elements $H_{ij}$. Then the above
property of off-diagonal entries implies the possibility of hopping throughout the
entire chain which results in the delocalized character of eigenvectors. If on the contrary
the off-diagonal matrix elements $H_{ij}$ are nonzero only inside the band $ \, |i-j| <
\lambda $, all the eigenvectors turn out to be localized in the same way as in a quasi
one-dimensional disordered wire \cite{MF91}. This {\it banded
random matrix} (BRM) theory is also relevant for chaotic systems under a
time-periodic perturbation \cite{KickRot}.

One well--known extension of the Wigner-Dyson RMT is the model of Rosenzweig and
Porter \cite{RosPort} (RPRM) with the variance $ \, \langle | H_{i \ne j} |^2 \rangle
 = {\cal B}^2 / N^{2\alpha} \, $ that does not depend on the distance $ \, | i -
j | \, $ but depends on the matrix size. The spectral statistics
of RPRM at different parameter $ \, \alpha \, $ can range from
almost uncorrelated levels (if $ \, \alpha > 1 \, $) to a strong
level repulsion which is similar to a weakly perturbed
Wigner-Dyson ensemble (if $ \, \alpha = 1/2 \, $), see, for
instance, the papers [\onlinecite{DatKunz, PandBrezHik,Guhr}] and
references therein. Exact calculation of the two--levels
correlation function of RPRM with the complex hopping entries $ \,
H_{i \ne j} \, $ and $ \, \alpha = 1 \, $ shows that the spectral
statistics falls neither into the Poissonian nor into the Wigner-Dyson
universality classes \cite{ShapKunz}. It is often called ``the
regime of crossover''. Much less is known about RPRM with either
real or quaternionic hopping elements \cite{DatKunz}.

Without lost of generality, RPRM can be presented by an ensemble
of the matrices $ \, \hat{H}_{RP} = \hat{A} + ({\cal
B}/N^{\alpha}) \hat{B} \, $ with a superposition of a diagonal
matrix $ \, \hat{A} \, $ and a full Wigner-Dyson matrix $ \,
\hat{B} \, $ of any symmetry ($\beta = 1, 2, 4$). If $ \, \hat{A}
\, $ is the random matrix possessing the Poissonian level
statistics and $ \, {\cal B} = 0 \, $ the model describes spectral
properties of a classical integrable system. By increasing $ \,
{\cal B} \, $ one can explore a transition from integrable to a
classically chaotic system with or without the time--reversal
symmetry \cite{Guhr}.

Recently another Gaussian RMT that interpolates between the Wigner-Dyson RMT and
BRM theory attracted a considerable attention. This is the {\it power law banded
random matrix} (PLBRM) theory \cite{MF,KM,ME,KrTs} for which $\langle |H_{ij}|^{2}
\rangle$ is nearly constant inside the band $ |i-j| < \lambda $ and decreases as a
power-law function $\langle |H_{ij}|^{2}\rangle \sim 1/|i-j|^{-2\alpha}$ for $
|i-j|> \lambda $. The special case $\alpha=1$ is relevant for description of critical
systems with multifractal eigenstates \cite{MF,KM,ME,KrTs,AltLev,Levitov}, in particular
for systems at the Anderson localization-delocalization transition point. The case
$\alpha >1$ corresponds to the power-law localization which can be found in certain
periodically driven quantum-mechanical systems \cite{KickRotPLRM}.

The progress in BRM and PLBRM theories  became possible because of
mapping \cite{MF91,MF} onto the nonlinear supersymmetric sigma-model \cite{Efetov}
that allowed to obtain rigorous results by using various powerful methods of the field
theory. However, such mapping is only justified if the bandwidth $ \lambda\gg 1$.
In the opposite case where all the off-diagonal matrix elements are
parametrically small compared to the diagonal ones, no field-theoretical approach
is known so far. Yet such
{\it almost diagonal} RMT may possess nontrivial properties because of the slow
decay of the off-diagonal matrix elements $\langle |H_{ij}|^{2}\rangle$ with
increasing $|i-j|$. For instance
it is  of fundamental interest to study the spectral statistics in systems
with {\it power-law localization} that takes place in the power-law banded random
matrix ensembles at $\alpha>1$.
Another problem to study is the critical almost diagonal PLBRM.
It is known that the eigenvectors of PLBRM with $\alpha=1$ remain multifractal
for an arbitrary small value of $ \, \lambda \, $ [\onlinecite{Levitov}].
This means that the typical eigenfunction is {\it extended} though very sparse at
small $ \, \lambda $. Thus almost diagonal random matrices may display the {\it
localization-delocalization transition} with changing the exponent
$ \, \alpha \, $ as well as their large bandwidth counterpart \cite{MF}. This
transition has been studied numerically \cite{PLRMnum} for $ \, \lambda \sim 1 \, $
but little is known about it in the limit of almost diagonal random matrices.

The goal of this paper is to develop  a formalism that would allow
to describe the spectral statistics for a {\it wide class of
almost diagonal random matrices} with a generic behaviour of $ \,
\langle |H_{i \ne j}|^{2}\rangle = b^{2}{\cal F}(|i-j|) \, $ and
$\langle|H_{ii}|^{2}\rangle \sim 1$. The parameter $ \, b \, $ may
or may not depend on the matrix size $ \, N $. The principle
requirement is that $ \, b \ll 1 \, $ is {\it small}. It controls
the smallness of the off-diagonal matrix elements and will be used
as an expansion parameter for the spectral correlation functions.

A natural way to proceed with such an expansion is to develop a perturbation
theory in the parameter $ \, b \, $ starting from the diagonal matrix as a
zero-order approximation. However, a naive expansion of the spectral form-factor
$ \, K(\tau) \, $ up to a finite order in $ \, b \, $ may fail, since the parameter
$ \, b \, $ enters in the product $ \, b N \tau \, $ with the matrix size $ \, N \, $
and the time $ \, \tau \, $ measured in units of the Heisenberg time. In this combination,
the small parameter $ \, b \, $ can be compensated by the (large) product $ \,  N \tau $.
In order to be able to obtain the {\it regular expansion in powers of $ \, b $}, $ \, K(\tau) =
1 + b K_{1} (\tau)+ b^{2} K_{2}(\tau)+ \ldots $, in the thermodynamic limit $ \, N \to
\infty \, $ at arbitrary $ \, N \tau $, we develop a diagrammatic technique that gives
$ \, K_m(\tau) \, $ in a form of {\it infinite series} in powers of $ \, b N \tau $.
It is essentially a kind of locator expansion \cite{Zim} adjusted to the problem of
spectral statistics. It formalizes and generalizes the idea of
Ref.[\onlinecite{Levitov}] where the resonance pairs of states $i,j$ with
$|H_{ii}-H_{jj}|< |H_{ij}|$ have been considered to find the statistics of
multifractal eigenstates. In our approach, the coefficients $ \, K_{m}(\tau) \, $ result
from an interaction (via the off-diagonal elements $H_{ij}$) of $ \, m+1 \, $ energy levels.
We call this expansion a {\it virial expansion} by analogy with the expansion of
thermodynamic functions of a dilute system in powers of density with {\it virial
coefficients} that take into account collisions of $ \, m + 1 \, $ particles. We stress that
what we are doing below is not an {\it approximation} of two, three, or more
{\it resonant} levels. It is rather the {\it classification} of exact
perturbation series by a number of the interacting levels involved.
The question of a contribution from non-resonant levels does not arise,
since at a given number of the interacting levels all possible relations between $ \,
|H_{ii}-H_{jj}| \, $ and $ \, |H_{ij}| \, $ are taken into account.

The most serious problems on that route are i) dealing with non-commutative matrices
$ \, \hat{H}_{\varepsilon}=diag\{H_{ii}\} \, $ and $ \, \hat{V}=\hat{H}-\hat{H}_{
\varepsilon} $; \ ii) combinatorial coefficients in the infinite series in $ \, b N \tau $.
The first problem is solved by using the {\it Trotter formula} (see  Section \ref{method}). The
second problem reduces to a particular case of the general problem of graph coloring that is
amenable to an exact solution (see Section \ref{2ColSec} and \ref{3ColSec}).

As a result we present the series for $K_{m}(\tau)$ with combinatorial coefficients that do
not depend on the particular choice of ${\cal F}(|i-j|)$ in the
variance $\langle |H_{ij}|^{2}\rangle = b^{2} {\cal F}(|i-j|)$. They can thus be
applied to {\it any} Gaussian ensemble of almost diagonal random matrices. The
particularly simple expressions are found for  $m=1,2$ which correspond to
two and three interacting levels.

The general formulae obtained in this way are illustrated in Section \ref{exmplRP} for
RPRM with $ \, b = {\cal B}/N \, ; \ {\cal F}(|i-j|) = 1 \, $ and in Section \ref{exmplPLBRM}
for a critical PLBRM with $ \, b = {\cal B} \, ; \ {\cal F}(|i-j|) \sim 1/|i-j|^{2}$. In
both examples, the constant $ \, {\cal B} \, $ will be taken small: $ \, {\cal B} \ll 1 $. As a
result, the correction to the Poisson behaviour $ \, K(\tau)=1 $ has been obtained in the limit
of infinite matrix size $ \, N \rightarrow \infty $. The existence of such a correction
can be interpreted as a rigorous detection of {\it delocalization} in the framework
of a formalism that starts from the diagonal random matrix where all states are
{\it localized}. Thus our theory offers a {\it controllable way of obtaining
delocalization from localization}. This is just complementary to conventional
approaches based on the  supersymmetric sigma-model \cite{Efetov} that
always starts from delocalized (e.g.diffusive) modes.

\section{Basic definitions}

Let us consider a Hermitian RM of size $ \, N \times N \, , \ N \gg 1 $, from a
Gaussian ensemble. We
assume that entries of the matrix are random and independent. The RM is the
Hamiltonian
$ \, \hat H \, $ of the matrix Schr{\"o}dinger equation
\[
   \hat H \psi_n = \epsilon_n \psi_n
\]
where $ \, \epsilon_n \mbox{ and } \psi_n \, $ are the eigenvalues and eigenvectors,
respectively. Define statistical properties of the matrix entries:
\begin{equation}\label{model}
   \langle H_{i,j}   \rangle = 0 \, ; \quad
   \langle H_{i,i}^2 \rangle = { 1 \over \beta } \, ; \qquad
   \langle | H_{i,j} |^2 \rangle = b^2 \, {\cal F}( | i - j | ) \, , \ i \ne j \, ;
\end{equation}
where $ \, {\cal F}(|i-j|) > 0 \, $ is a smooth function of its argument, and the
parameter $ \, b \, $ is small:
\[
   b \ll 1 \, .
\]
The condition $ \, b \ll 1 \, $ means that RM is {\it almost diagonal}. The
parameter $
\, \beta \, $ corresponds
to the Dyson symmetry classes: $ \, \beta_{GOE} = 1 $, $ \ \beta_{GUE} = 2 $. The
brackets $ \, \langle
\ldots \rangle \, $ denote the ensemble averaging. If the ensemble is Gaussian the
averaging
of a function $ \, F(A) \, $ over a random variable $ \, A \, $ reads:
\[
   \langle F(A) \rangle_A \equiv
   { 1 \over \sqrt{ 2 \pi \bar A^2 } } \ \int_{-\infty}^{+\infty} F(A) \exp \left( -
         \frac{A^2}{2\bar A^2} \right) \, {\rm d} A \, .
\]

We concentrate on the level statistics which is characterized by the density of
states $ \, \rho(E) =
\sum_n \delta( E - \epsilon_n ) $ and its multi-point correlation functions.
For example, the two-level
correlation function $ \, R(\omega) \, $ is defined as:
\begin{equation}
\label{R-corr}
     R(\omega) = { \langle \langle \, \rho(\omega/2) \, \rho(-\omega/2) \, \rangle
\rangle \over \langle \,
                     \rho(0) \, \rangle^{2} } \, ; \quad
     \langle \langle \, \hat{a} \, \hat{b} \, \rangle \rangle \equiv \langle \,
\hat{a} \,
\hat{b} \,
            \rangle - \langle \, \hat{a} \, \rangle \langle \, \hat{b} \, \rangle \, .
\end{equation}
The Fourier transform of $ \, R(\omega) \, $ is known as the
spectral form-factor~$ \, K(t) $:
\begin{equation}
   K(t) = \int_{-\infty}^{+\infty} e^{{\bf i}\,\omega t} \, R( \omega ) \,
{\rm d} \omega \, .
\end{equation}
We will see below that for almost diagonal RM the representation of spectral
statistics in the time domain turns out to be convenient.
Therefore from now on we use the form factor instead of the correlation function.
We rescale time by the mean level spacing $ \, \Delta \equiv 1 / \langle \, \rho(0) \,
\rangle $ introducing the dimensionless time $ \, \tau = t \, \Delta $. Our goal is to
develop a regular perturbative expansion for $ \, K (\tau) \, $ in powers of small
parameter $ \, b \ll 1 $. In the limit of small time the spectral form-factor
$ \, K( \tau \to 0 ) \, $ is linked to the other important spectral characteristics
called {\it the level compressibility} \cite{CKL}:
\begin{equation}\label{chi}
   \chi = \lim_{\tau \to 0} \Bigl( \lim_{N \to \infty} K( \tau ) \Bigr) \, .
\end{equation}
Let us take a window of the width $ \, \delta E $, $ \, \delta E / \Delta \equiv \bar{n}
\ll N $, in the energy space centered at $E=0$ and calculate the number $ \, n \, $
of levels inside the window at some realization of disorder.
The level number variance is $ \,
\Sigma_2( \bar{n} ) = \langle ( n - \bar{n} )^2 \rangle $.
The level compressibility is by definition
the limit
\begin{equation}\label{chi-def}
   \chi = \lim_{ \bar{n} \to\infty} \left( \lim_{N\to\infty}
{ \partial \, \Sigma_2 \, ( \bar{n} )
                     \over \partial \, \bar{n} } \right) \, .
\end{equation}
The level compressibility contains an information about the localization transition:
$ \, \chi \, $ ranges
from $ \, \chi_{WD} = 0 \, $ for the Wigner-Dyson statistics with
extended wave functions and a strong levels repulsion to $ \,
\chi_{P} = 1 \, $ in the case of localized wave functions and uncorrelated
levels with the Poissonian distribution. The intermediate situation with
$ \, 0 < \chi_{crit} < 1 \, $ is inherent for the critical regime of multifractal
wave functions \cite{CKL}.

The main object of our further analysis is the following correlation function in the
time domain:
\begin{equation}
\label{K0}
\tilde{K}_{N}(\tau)=\displaystyle \frac{1}{N}
            \langle\langle \, {\rm Tr} \, e^{ -{\bf i} \, \hat{H} \, \tau / \Delta } \,
                              {\rm Tr} \, e^{  {\bf i} \, \hat{H} \, \tau / \Delta }
            \, \rangle\rangle \equiv \tilde K_0(\tau) + b \tilde K_1(\tau) +
                                         b^2 \tilde K_2(\tau) + \ldots
\end{equation}
For the constant mean density of states $\tilde{K}(\tau)$ coincides with $K(\tau)$.
However, they are different if $\langle \rho(E)\rangle$ essentially depends on energy
$E$. In analogy with Eq.(\ref{chi}) one can define the quantity
$\chi_0 \equiv \lim_{\tau \to 0} \Bigl( \, \lim_{N \to \infty} \tilde{K}_{N}(\tau) \Bigr)$.
It turns out that at small $ \, b \, $ there is a simple approximate relationship between
$ \, \chi \, $ and $ \, \chi_0 \, $ (see Appendix \ref{AppDef}):
\begin{equation} \label{K-chi}
\chi \simeq 1 - \displaystyle { 1 - \chi_0 \over \Upsilon }, \quad \Upsilon =
\displaystyle \displaystyle\frac{\Delta}{N} \int_{-\infty}^{+\infty}
                       \langle \rho(E) \, \rangle^2 \, {\rm d} E.
\end{equation}
The mean density of states for the Gaussian ensemble of almost
diagonal RMs with either localized or (sparse) fractal eigenstates
is close to the Gaussian distribution of the diagonal entries
\cite{KrMac}:
\begin{equation}\label{scales}
   \langle \rho(E) \, \rangle \simeq N \sqrt{ \beta \over 2 \pi } \exp
\left( - \frac{\beta E^2}{2} \right)
      \ \Rightarrow \
   \Delta \simeq {1 \over N} \sqrt{ 2 \pi \over \beta } .
\end{equation}
Thus with an accuracy of $ \, O(b) $ the unfolding factor in Eq.(\ref{K-chi}) 
can be taken as $ \, \Upsilon^{-1} \simeq \sqrt{2} $.

\section{The method}\label{method}

\subsection{The Trotter formula}

As far as we investigate the properties of almost diagonal RMs the Hamiltonian can
be naturally
divided into a diagonal part $ \, \hat{H}_\varepsilon \, $ and a matrix of hopping
elements $ \, \hat{V} $:
\begin{equation}\label{sep}
  \hat{H} \equiv \hat{H}_\varepsilon + \hat{V} \, .
\end{equation}
It follows from the definition (\ref{model}) that the hopping elements  $ \,
H_{i,j} \equiv V_{i,j} \sim b \, $ are small compared to the diagonal ones
$ \, H_{i,i} \equiv \varepsilon_i \sim 1 $. However, matrices
$ \, \hat{H}_\varepsilon \mbox{ and } \hat{V} \, $ do not commute with each
other and a direct expansion of the exponentials $ \, e^{ \pm {\bf i} \, (
\hat{H}_\varepsilon + \hat{V} ) { \tau \over \Delta } } \, $ in Eq.(\ref{K-chi})
in terms of $ \, \hat{V} \, $ involves serious difficulties.
One possible way to overcome these problems is to represent $ \, e^{ \pm {\bf i} \,
(
\hat{H}_\varepsilon + \hat{V} ) { \tau \over \Delta } } \, $
as a product of exponential functions containing matrices $ \,
\hat{H}_\varepsilon \mbox{ and } \hat{V} \, $ separately.
To do this we use a generalization of the identity $ \, \exp ( a + b ) \equiv
\exp( a ) \exp( b ) \, $ for non-commuting variables
known as the {\it the Trotter formula} \cite{Trt}:

\begin{equation}\label{Trott}
  e^{\hat{A}+\hat{B}} = \lim_{n\to\infty} \left( e^{\hat{A}/n} \, e^{\hat{B}/n} \right)^n
       \quad \Rightarrow \quad
   e^{ \pm {\bf i} \, \hat{H} { \tau \over \Delta } } =
\lim_{n\to\infty}\prod_{p=1}^{n}
               \left(  e^{ \pm {\bf i} \, \hat{H}_\varepsilon { \tau \over \Delta n} }
                       e^{ \pm {\bf i} \, \hat{V}          { \tau \over \Delta n} }
\right) \, .
\end{equation}
The Trotter formula is exact for the finite matrices
$ \, \hat{A} \mbox{ and } \hat{B} $. Therefore, correct order of limits reads:
\begin{equation}\label{CorrLim}
   \left\{
     \begin{array}{l}
       1. \ n \to \infty \, ; \cr
       2. \ N \to \infty \, ; \cr
       3. ({\rm if}\;\;  {\rm necessary}) \ \tau \to 0 \, .
     \end{array}
   \right.
\end{equation}
In what follows we will always imply precisely this order of doing limits.

Eq.(\ref{Trott}) allows to make a regular expansion in powers $ \, \hat{V} $.
The price for that is the infinite product in Eq.(\ref{Trott}). We need a
proper selection rule to extract from that product terms of the order of
$ \, O(b^0), \, O(b^1), \, O(b^2), \ldots $.

\subsection{Lowest-order terms}
For a strictly diagonal matrix the spectral form-factor
$\tilde{K}(\tau)$ is calculated straightforwardly:
\begin{equation}\label{Pcalc}
  \tilde{K}(\tau)|_{\hat{V}=0}={ 1 \over N }
  \langle\langle
  {\rm Tr} \, e^{ - {\bf i} \, \hat{H}_{\varepsilon} \tau/\Delta } \ {\rm Tr} \, e^{
{\bf i} \,
\hat{H}_{\varepsilon} \tau/\Delta }\rangle\rangle = \frac{1}{N}\sum_{l}
\left[1-\langle
e^{-i\varepsilon_{l}\tau/\Delta}\rangle
^{2}\right]=1-e^{-\langle\varepsilon^{2}\rangle (\tau/\Delta)^{2}}
\end{equation}
For all Gaussian RM of the type Eq.(\ref{model}) the inverse mean level spacing
increases with the matrix size $N$. The inverse mean level spacing $ \, 1/\Delta \equiv
\tilde{N} \, $ is proportional to the matrix size: $ \, \tilde{N} \sim N $. For finite $ \,
N \, $ the normalization sum rule \cite{CKL} requires $ \, \tilde{K}(0)=0 $. If however the
limit $ \, N\rightarrow\infty \, $ is done {\it prior} to the limit $ \, \tau
\rightarrow 0 \, $ (see Eq.(\ref{CorrLim})) the normalization sum rule is violated and
we recover the Poisson statistics result $ \, \tilde{K}(\tau)=1 $.

In order to obtain  corrections to $\tilde{K}(\tau)$ proportional to $b^{k}$
one has to expand $e^{\pm i \tilde{N}\tau \hat{V}/n}$ in powers of $\hat{V}$
in the infinite product in the r.h.s. of Eq.(\ref{Trott}):
\begin{equation}\label{chain-1}
    \underbrace{ \ldots
           \times
           \exp \left( \pm {\bf i} { \tilde{N}\tau \over n }
\hat{H}_{\varepsilon } \right) \,
           \exp \left( \pm {\bf i} { \tilde{N} \tau \over n } \hat{V}
\right) \times
           \exp \left( \pm {\bf i} { \tilde{N} \tau \over n } \hat{H}_{\varepsilon }
\right) \,
           \exp \left( \pm {\bf i} { \tilde{N} \tau \over  n } \hat{V}
\right)
\times
                 \ldots }_{ \mbox{ {\it n} pairs of exponentials } } \, .
\end{equation}
and then to perform the Gaussian averaging over $\hat{H}_{\varepsilon}$ and $\hat{V}$.
Note that the Gaussian average is zero for all terms in r.h.s of Eq.(\ref{K0}) which
contain an odd number of off-diagonal matrices $ \, \hat{V} $. That is why only terms
with {\it even} powers of $ \, b \, $ may result from such an expansion. For the
purpose of explaining the details of the new formalism based on the Trotter formula we
show how the term $ \, \propto b^{2} \, $ arises from it.

\subsubsection{Gaussian averaging}
Each $ \, {\rm Tr}\,  e^{\pm i H \tilde{N}\tau} \, $ in Eq.(\ref{K0}) can be represented
using the Trotter formula as ${\rm Tr}$ of the infinite product Eq.(\ref{chain-1}).
Therefore we have to expand either {\it two} exponentials $e^{\pm i
\hat{V}\tilde{N}\tau/n}$ in Eq.(\ref{chain-1}) up to $\hat{V}$ or one single
exponential up to $\hat{V}^{2}$ setting $\hat{V}=0$ in all the other exponentials $ \,
e^{\pm i \hat{V}\tilde{N}\tau/n} $. We will show below that only the first option survives
the limit $n\rightarrow\infty$.

Expanding two of $ \, \exp \left( \pm {\bf i} { \tilde{N} \tau \over n } \hat{V}
\right) \, $ in Eq.(\ref{chain-1}) up to $ \, \hat{V}^1 $ we obtain \cite{Loc}:
\begin{equation}\label{chain}
           \exp \left( \pm {\bf i} { \tilde{N} \tau p_1 \over n }
\hat{H}_{\varepsilon }
\right) \,
           \left( \pm {\bf i} { \tilde{N} \tau \over n } \hat{V}
\right)
           \exp \left( \pm {\bf i} { \tilde{N} \tau p_2 \over n }
\hat{H}_{\varepsilon }
\right) \,
           \left( \pm {\bf i} { \tilde{N} \tau \over n } \hat{V}
\right)
           \exp \left( \pm {\bf i} { \tilde{N} \tau p_3 \over n }
\hat{H}_{\varepsilon }
\right) \sim
           { b^2 } \, ,
\end{equation}
where $ \, p_{1,2,3} \, $ are the numbers of successive exponentials $ \, \exp \left(
\pm {\bf i} { \tilde{N} \tau  \over n } \hat{H}_{\varepsilon } \right) \, $ `fused'
together after we set $\hat{V}=0$ in all $ \, \exp \left( \pm {\bf i} { \tilde{N}
\tau \over n } \hat{V} \right) \, $ but two. We call
$ \, p_{s} \, $ {\it the Trotter numbers}. They obey the obvious restriction
\begin{equation}\label{norm}
      p_1 + p_2 +p_3 = n \, , \quad p_{1,2} \ge 1 \, , \ p_3 \ge 0 \, .
\end{equation}

Note that both off-diagonal matrices $ \, \hat{V} \, $ in Eq.(\ref{chain}) must belong
to the same trace. If they belong to the different traces the result is zero because
$ \, {\rm Tr} \, \Bigl\{ \exp \left( \pm {\bf i} { \tilde{N} \tau p_i \over n }
\hat{H}_{\varepsilon } \right) \hat{V} \Bigr\} = 0 $.

In order to find the term $ \, {\cal C} \propto O(b^2) $ in Eq.(\ref{K0})  we
substitute Eq.(\ref{chain}) into Eq.(\ref{K0})
and perform the Gaussian averaging in accordance with Eq.(\ref{model}):
\begin{eqnarray}
   {\cal C} =
   - { \frac{4}{N}}  \left( { \tilde{N} \tau \over n } \right)^2
   \sum_{\{p_s\}}^n \langle \langle
     \sum_{i,j,m=1}^N
        e^{ - {\bf i} { \tilde{N} \tau p_1 \over n } \varepsilon_i }
          \, V_{i,j} \,
        e^{ - {\bf i} { \tilde{N} \tau p_2 \over n } \varepsilon_j }
          \, V_{j,i} \,
        e^{ - {\bf i} { \tilde{N} \tau p_3 \over n } \varepsilon_i } \times
        e^{ {\bf i} \tilde{N} \tau \varepsilon_m }
   \rangle \rangle |_{m=i \mbox{ or } m=j}
   \label{b2Transf}
\end{eqnarray}
\[
   = - 4 { \left( b \tilde{N} \tau \right)^2 \over n^2 } \, {\cal R}_N(1)
     \sum_{\{p_s\}}^n \left\{
       \, {\cal D}_N \left( { n - p_1 - p_3 \over n } \right)
          {\cal D}_N \left( { p_2 \over n } \right) +
       \, {\cal D}_N \left( { p_1 + p_3 \over n } \right)
          {\cal D}_N \left( { n - p_2 \over n } \right)
     \right\} \, ;
\]
where $ \, \displaystyle \sum_{\{p_s\}}^n \, $ means summation over the Trotter numbers
$ \, \displaystyle \sum_{p_1,p_2, p_3 = 1}^n \!\!\! \delta_{n-(p_1+p_2+p_3)} \, $ and
we denote
\begin{equation}\label{DandR}
      {\cal D}_N (y) \equiv \exp \left[
                            - { 1 \over 2 \beta } \left( \tilde{N} \tau \right)^2 y^2
                             \right] \, ;
  \quad
     {\cal R}_N (k) \equiv { 1 \over N } \,
       \sum_{i>j}^N \Bigl(  {\cal F}(|i-j|) \, \Bigr)^k \, .
\end{equation}

\begin{figure}
\unitlength1cm
\begin{picture}(15,4.25)
   \epsfig{file=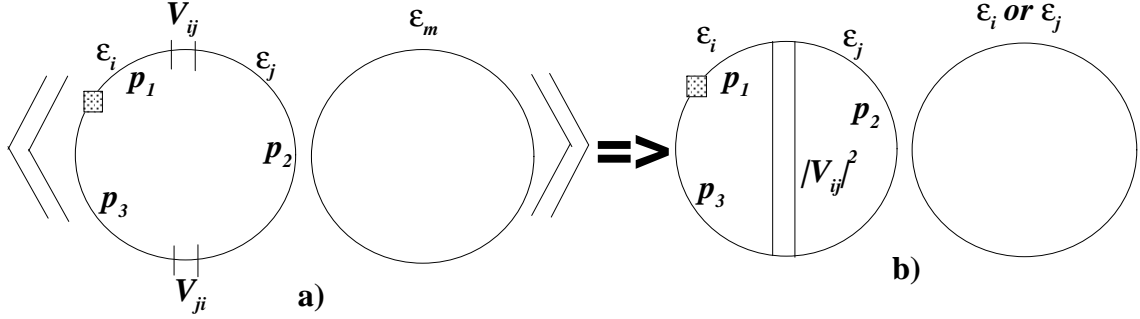,angle=0,width=15cm}
\end{picture}
\vspace{0.5cm}
\caption{
\label{Diag1}
Graphic illustration of  Eq.(\ref{b2Transf}) before (a) and after (b) averaging.
}
\end{figure}

\noindent
The function $ \, {\cal R}_N (k) \, $ depends on the particular form
of decay of the off-diagonal matrix elements in the Gaussian ensemble
(see Eq.(\ref{model})) and contains the sum over matrix indices which
will be referred to below  as the summation over the `real space'.

It is instructive to give a graphic representation of Eq.(\ref{b2Transf}) before and
after averaging. In
Fig.\ref{Diag1} each circle denotes the trace. The vertices $ \, V_{i,j} \mbox{ and
}
V_{j,i} \, $ in the left circle are uncoupled before averaging (Fig.\ref{Diag1}a).
The shadowed box indicates starting point of the Trotter chain Eq.(\ref{chain-1}).
The circle is thus divided into
segments (below referred to as the {\it Trotter segments}) corresponding to $ \,
e^{ - {\bf i} { \tilde{N} \tau p_1 \over n } \varepsilon_i } \, , \ e^{ - {\bf i} {
\tilde{N} \tau p_2 \over n }
\varepsilon_j } \, , \mbox{ and } e^{ - {\bf i} { \tilde{N} \tau p_3 \over n }
\varepsilon_i } $. We mark the
length of each segment by the Trotter number $p_{s}$. The right circle stands for $
\, e^{ {\bf i} \tilde{N} \tau
\varepsilon_m } $. The averaging over $\hat{V}$ connects uncoupled $V$-vertices by the
correlation function $ \, | V_{ij} |^2 =
b^{2} {\cal F}(|i-j|)$ denoted by a double line while the averaging over
$\varepsilon_{m}$ couples  $ \, \varepsilon_m \, $
to either $ \, \varepsilon_i \, $ or $ \, \varepsilon_j \, $ (Fig.\ref{Diag1}b).

\subsubsection{Integral over Trotter variables}\label{IntRig}

One has to sum over the Trotter numbers in Eq.(\ref{b2Transf}) prior to doing the limit
$ \, n \to \infty $. To this end we introduce the
{\it Trotter variables} $y_{s}=p_{s}/n$ which in the limit $n\rightarrow\infty$
can be considered as continuous. Then the summation over the Trotter numbers $p_{s}$
can be replaced by integration over the Trotter variables. Switching from the
Kronecker-delta  $ \, \delta_p \, $ to the Dirac $\delta$-function
$n\delta_{p}\rightarrow\delta(y)$ we arrive at:
\begin{eqnarray}\label{TrI}
  {\cal C} & = & - 4 \left( b \tilde{N} \tau \right)^2 \, {\cal R}_N(1) \,
                      {\cal I} \, ; \cr
  \displaystyle
   {\cal I} & = &
   \int \!\! \int \!\! \int_0^1 {\rm d}y_1 \, {\rm d}y_2 \, {\rm d}y_3 \,
     \delta( 1 - y_1 - y_2 - y_3 )
     \Bigl\{
       \, {\cal D}_N \left( 1 - y_1 - y_3 \right)
          {\cal D}_N \left( y_2 \right) + \\
& \ &  \qquad   + {\cal D}_N \left( y_1 + y_3 \right)
                  {\cal D}_N \left( 1 - y_2 \right)
     \Bigr\} \, .
  \nonumber
\end{eqnarray}
It is remarkable that as a result of this transformation, which is exact in
$n\rightarrow\infty$ limit, all the $n$-dependent factors are
absorbed by $y_{s}$. This is because the number of independent Trotter
numbers $p_{s}$ in Eq.(\ref{b2Transf}) coincides with the number of matrices
$\hat{V}$, each coming with the factor $1/n$. Note that this is not the case
if only one exponential $e^{\pm i \hat{V}\tilde{N}\tau/n}$ in Eq.(\ref{chain-1}) is
expanded up to $\hat{V}^{2}$ order. Then the number of Trotter summations is less by
one which leaves an uncompensated factor $ \, 1/n \rightarrow 0 \, $
after switching to the Trotter variables. Thus we arrive at an important conclusion
that {\it in all orders} of $\hat{V}$-expansion one should retain only {\it
linear} terms of expansion of the exponential $ \, e^{\pm i \hat{V}\tilde{N}\tau/n}
\approx 1 \pm i \hat{V}\tilde{N}\tau/n $ [\onlinecite{PathInt}].

The integral Eq.(\ref{TrI}), which will be referred to as `the Trotter integral',
can be simplified by observing that the $\delta$-function
imposes the same arguments in both ${\cal D}_{N}$  functions in the corresponding
products. Thus $ \, {\cal D}_{N} \, $ functions can be fused $ \, {\cal D}_{N}(x) \,
{\cal D}_{N}(x) \equiv {\cal D}_N \left( \sqrt{2} \, x \right) $ and the Trotter
integral takes the form:
\begin{eqnarray}\label{TrI-2}
 {\cal I} = \int_0^1 {\rm d} y \, ( 1 - y )
     \Bigl\{
       \,
          {\cal D}_N \left( \sqrt{2} \, y \right) +
       \,
          {\cal D}_N \Bigl( \sqrt{2} \, [ 1 - y ] \Bigr)
     \Bigr\} \, .
\end{eqnarray}
The similar fusion takes place in the Trotter integrals at an arbitrary order of
$V$-expansion.

The further simplification is possible only in doing the thermodynamic limit
$N\rightarrow\infty$. In this limit the function ${\cal D}_{N}(y)$ defined in
Eq.(\ref{DandR}) collapses to the $\delta$-function:
\begin{equation}\label{delta}
   {\cal D}_N (\sqrt{2}y) |_{N\to\infty} \simeq
\frac{\sqrt{\pi\beta}}{\tilde{N}|\tau|}\;\delta(y).
\end{equation}
Then the integral (\ref{TrI-2}) is easily calculated and we find finally the
expression for $ \, {\cal C} $:
\begin{equation}\label{Term1}
  {\cal C} =  - b \, \sqrt{ \beta \pi } \, \left( b \tilde{N} |\tau| \right) \,
             {\cal R}_N(1) \, ,
\end{equation}
where $ \, {\cal R}_N(1) \, $ is given by Eq.(\ref{DandR}).

\begin{figure}
\unitlength1cm
\begin{picture}(4.5,1.25)
   \epsfig{file=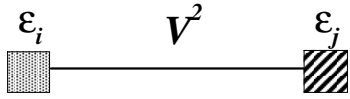,angle=0,width=4.5cm}
\end{picture}
\vspace{0.25cm}
\caption{
\label{Diag1-1}
Graphic illustration of the simplest term $ \, {\cal C} $.
}
\end{figure}

Let us look at the calculation of ${\cal C}$ from a different viewpoint. Namely,
we note that in this calculation only two species of levels $ \, i \, $ and
$ \, j \, $ are involved. We show this situation graphically
in~Fig(\ref{Diag1-1}).
Shadowed boxes mark the energy levels with different shadowing (``colors'')
for  $ \, \varepsilon_i \, $ and
$ \, \varepsilon_j $. They are connected by the interaction line (IL)
which is associated with the factor $ \, ( \tilde{N} \tau b )^{2} $.
Because of the fusion of two ${\cal D}_{N}$-functions,  the interaction of {\it two}
levels brings only {\it one} $\delta$-function with the normalization factor
$1/(\tilde{N}|\tau|)$. As the result, we get one `free' parameter $b$ which
is decoupled from the combination $ \, \tilde{N} |\tau| \, $ that involves
the matrix size $ \, N $:
\[
   {\cal C} \sim { ( \tilde{N} \tau b)^2 \over \tilde{N} |\tau| } {\cal R}_N(1) =
               b^1 \times ( \tilde{N} |\tau| b)^1 \ {\cal R}_N(1) \, .
\]
It is clear from the above analysis that if we continue expanding in higher powers of
$ \, V_{ij}=V_{ji}^{*} \, $ but keep the number of colors equal to 2, we will increase the
power of the combination $(b\tilde{N}|\tau|)$ but not the power of the free parameter $ \, b $.

\subsection{Selection rule}

Consider a generic term of order $b^{2k}$ in the perturbation series:
\begin{equation}\label{C-exp}
   {\cal C}_{2k} ({\sf A},{\sf B }) =
\, C_{N}(\{k_{i}\}) \,\, b^{\sf A} \ ( \tilde{N} |\tau| b)^{\sf B };
\quad {\sf A}+{\sf B } \equiv 2 k \, .
\end{equation}
It corresponds to a diagram with $ \, k = k_{1}+k_{2}+ \ldots \, $ interaction
lines distributed in a certain way $ \, \{ k_i \} = \{k_{1},k_{2}, \ldots \} \, $
among the links connecting a given number of different energy levels (``{\it colors}'').

Let us formulate the selection rule for such a perturbation theory.

\begin{figure}
\unitlength1cm
\begin{picture}(15,4.25)
   \epsfig{file=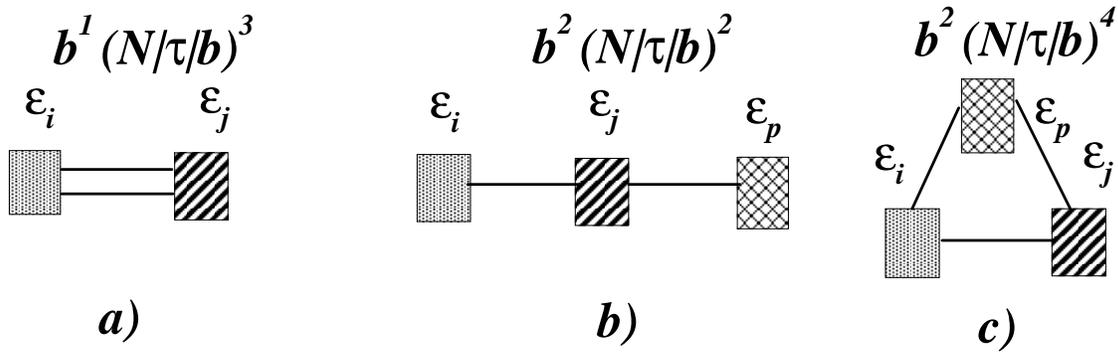,angle=0,width=15cm}
\end{picture}
\vspace{0.5cm}
\caption{
\label{SR}
Example of diagrams classified in accordance with the selection rule:
a) the case of two colors: $k=k_{1}=2$;
b) and c) two simplest diagrams for the case of three
colors: (b) $k=2,k_{1}=1,k_{2}=1$ and (c): $k=3, k_{1}=1,k_{2}=1,k_{3}=1$.
}
\end{figure}

\begin{itemize}

\item The power $ \, {\sf A} \, $ of the ``free'' parameter $ \, b \, $ is equal to
      the number of interacting energy levels minus 1: $ \, {\sf A} = c - 1 $.

\item The sum $ \, {\sf A}+{\sf B}  \, $ is equal to  $ \, 2k \, $.

\end{itemize}

We exemplify it in Fig.\ref{SR} for three different diagrams: the diagram with two colors
(a) corresponding to $ \, k = 2 \, , \ {\sf A} = 1 \, , \ {\sf B} = 3 $; and two diagrams
with three colors: (b) for $ \, k = 2 \, , \ {\sf A} = 2 \, , \ {\sf B} = 2 $; and (c) for
$ \, k = 3 \, , \ {\sf A} = 2 \, , \ {\sf B} = 4 $.

Suppose that we want to derive the perturbative term $ \, \tilde{K}_{\sf A}( \tau ) \, $ of
the order of $ \, O(b^{\sf A}) $, which would be valid at an arbitrary value of $ \, \tilde{N}
|\tau| b \, $ in the limit $ \, N \to \infty $. Such physical result can be obtained using the
following strategy:

\begin{enumerate}

\item Fix the power $ \, {\sf A} \, $ of the ``free'' small parameter $ \, b $ by
      fixing the number of the interacting energy levels, i.e., the number of colors:
      $ \, c $;

\item Perform summation of an infinite series in $ \, \tilde{N} |\tau| b \, $ at fixed $ \, c $
      over the number(s) $ \, k_{i} \, $ of IL along each link.

\item Analyze the limit $ \, N \to \infty $ of the corresponding {\it infinite} series in
      $ \, \tilde{N} |\tau| b$.

\end{enumerate}

\noindent
If this virial expansion works the series should converge in the limit $ \, N \to \infty $.

The strategy we suggested is a certain way of summation of
perturbative series. Its physical meaning is analogous to the idea
of resonant energy levels \cite{Levitov} or locator expansion \cite{Zim}.
However, in contrast to these essentially heuristic approaches
our selection rule offers a formalism that allows to implement a regular expansion
in the small parameter $ \, b $. It takes into account resonant as well as
non-resonant levels accurately which is impossible in the above heuristic methods.

Concluding this section we would like to note that the coefficient
$C_{N}(\{k_{i}\})$ in Eq.(\ref{C-exp}) is a product of two different factors:
\begin{equation}
\label{C-fac}
C_{N}(\{k_{i}\})={\cal R}_{N}(\{k_{i}\})\;{\cal V}(\{k_{i}\}),
\end{equation}
where $ \, {\cal V}(\{k_{i}\}) \, $ is a universal coefficient which depends on a
combinatorial factor $ \, {\cal K} \, $ and on the corresponding Trotter
integral $ \, {\cal I} $. The factor $ \, {\cal R}(\{k_{i}\}) \, $ is not
universal. It arises because of the summation of the product of
the correlation functions ${\cal F}(|i-j|)$ over the real
space and is a generalization of the function ${\cal R}_{N}(k)$ given by
Eq.(\ref{DandR}). In the large-$N$ limit it can be represented as:
\begin{equation} \label{1/N}
   {\cal R}_{N}(\{k_{i}\}) = {\cal R}(\{k_{i}\}) + \frac{1}{N} {\cal R}^{(1)}(\{k_{i}\}) + \ldots
\end{equation}
where $ \, {\cal R}(\{k_{i}\}) \, $ and $ \, {\cal R}^{(1)}(\{k_{i}\}) \, $ are of the same order.
In this paper, we study mainly a contribution of the leading term $ \, {\cal R}(\{k_{i}\}) $.
The role of the $ \, 1/N \, $ corrections to $ \, {\cal R}(\{k_{i}\}) \, $ is discussed below
in Section \ref{exmplPLBRM}.

\section{The case of two colors}\label{2ColSec}

In this section we calculate the term of order $ \, b \, $ in  $ \, \tilde{
K}(\tau) $. According to a general scheme outlined in the previous section this
correction is governed by the two-color diagrams:

\begin{figure}
\unitlength1cm
\begin{picture}(15,2.75)
   \epsfig{file=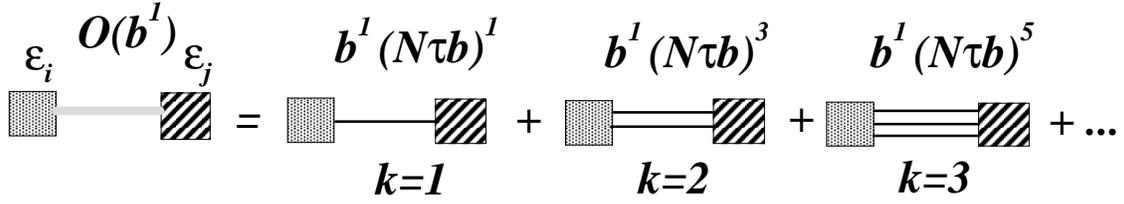,angle=0,width=15cm}
\end{picture}
\vspace{0.5cm}
\caption{
\label{2Col}
The two-color-diagrams.
}
\end{figure}

\noindent
which correspond to the following power series:
\begin{equation}\label{2ColSer}
   b \tilde{K}_1 =
   \sum_{k=1}^{\infty} (-1)^k \ x^{2k} \ {\cal V} (k) \ {\cal R}_N(k) \, ; \quad
       {\cal V} (k) \, =  \, {\cal K}(k) \ {\cal I}(k) \, ; \quad
       x \equiv \tilde{N} | \tau | b \, .
\end{equation}

The main difficulty in this calculation is the universal factor $ \, {\cal V}(k) \, $
in Eq.(\ref{2ColSer}). Applying the Trotter formula to both traces in Eq.(\ref{K0}) and
making an expansion in $ \, \hat{V} \, $ analogous to Eq.(\ref{chain}) we observe that
the same number of IL may correspond to completely different
distribution of the off-diagonal matrices $ \, \hat{V} \, $ over two traces (see Fig.5).
\begin{figure}[b]
\unitlength1cm
\begin{picture}(15,4.25)
   \epsfig{file=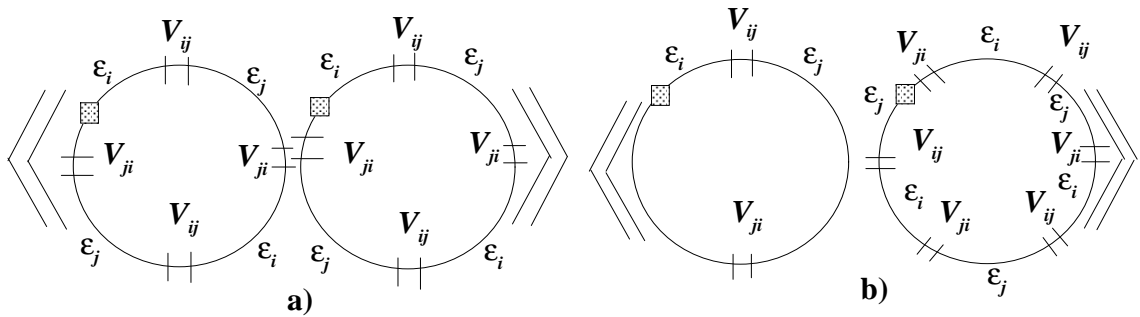,angle=0,width=15cm}
\end{picture}
\vspace{0.5cm}
\caption{
\label{Distr}
Two examples of different distribution of $\hat{V}$ at $ \, k = 4 $. The first
segment of the Trotter chain is denoted by the
shadowed box.
}
\end{figure}

The combinatorial part of the calculation consists of two tasks.
One of them is to distribute colors (denoting $\varepsilon_{i,j}$)
over $ \, 2k \, $ segments. Since $\hat{V}$ is the off-diagonal matrix,
the adjacent colors in each circle must be different. In general this is a
particular case of the famous combinatorial problem of graph coloring
\cite{gr}. For the case of only two colors ($\varepsilon_{i}$
and $\varepsilon_{j}$) considered in this section the problem is trivial:
each circle must contain an even number of the segments while
the first segment of the Trotter chain corresponds either to $\varepsilon_{i}$
or to $\varepsilon_{j}$ (see Fig.6), so that only the factor of 4 arises from
this coloring.
\begin{figure}[b]
\unitlength1cm
\begin{picture}(10,4)
   \epsfig{file=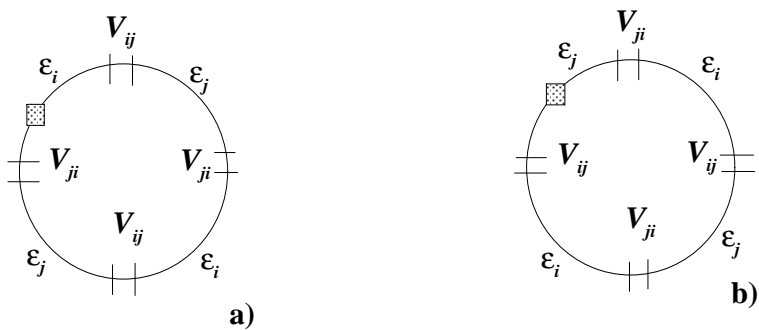,angle=0,width=10cm}
\end{picture}
\vspace{0.5cm}
\label{Color2C}
\caption{
Two different ways to put colors in the two-color problem. The first segment
is denoted by the shadowed box.
}
\end{figure}
The second combinatorial task is to find the number of ways to connect
$V$-matrices in pairs making either in-circle or
inter-circles connections. Such connections are equivalent to the Gaussian
averaging.
\begin{figure}[t]
\unitlength1cm
\begin{picture}(8,4)
   \epsfig{file=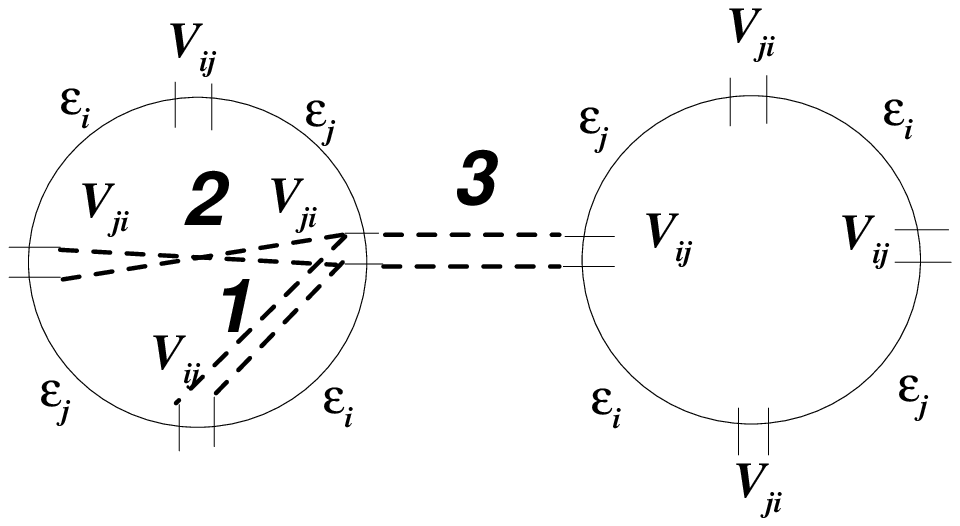,angle=0,width=8cm}
\end{picture}
\vspace{0.5cm}
\caption{
\label{Vcon2C}
An example for 3 different connections of $V$-vertices at $ \, k = 4 $.
}
\end{figure}

In Fig.7 we show three different ways to connect {\it two}
$V$-matrices. Two links labeled by 1 and 2 are in-circle connections, while the
third one (labeled by 3) is the inter-circles connection. The difference between
the connections 1 and 2 is that the connection 1 corresponds to the average
$\langle V_{ij}V_{ji} \rangle$ while the connection 2 corresponds to the average
$\langle V_{ij}V_{ij} \rangle$. For {\it real} Hermitian matrices
(the {\it orthogonal} ensemble, GOE, $\beta=1$) $V_{ij}=V_{ji}$
both connections are possible and the corresponding averages are equal to each
other. However, for {\it complex} Hermitian matrices with the same variance of the real
and imaginary parts (the {\it unitary} ensemble, GUE, $\beta=2$) only one of the
two averages, $ \, \langle V_{ij}V_{ji} \rangle $,  is non-zero. In this case the "crossed"
connection 2 is not allowed.

Now one can easily find the combinatorial factors associated with the
number of possible connections. In the GOE any of the $2k$ vertices $V$ can be
connected with any other $V$-vertex and we get:
\begin{equation}\label{GOEVcomb}
   {\cal K}_{GOE} (k) = 4 \, \times ( 2 k - 1 ) !! \, .
\end{equation}
In the GUE the $2k$ $V$-vertices are divided by two groups: one group containing
$k$ vertices of $V_{ij}$ and another group containing $k$ vertices $V_{ji}$.
The connection is possible only if $V$-vertices belong to {\it different} groups.
Thus we obtain for the corresponding combinatorial factor:
\begin{equation}\label{GUEVcomb}
   {\cal K}_{GUE} (k) = 4 \, \times k! \, .
\end{equation}

In order to find the coefficient $ \, {\cal I}(k) \, $ we have to fix the number of
$\hat{V}$ matrices in each trace to be $ \, 2m \, $ and $ \, 2( k-m) $, thereby also
fixing the number of the segments in the circles. Each segment corresponds to a Trotter
variable to be integrated over. The details of the calculation of the Trotter integral
$ \, {\cal I}(k) \, $ are presented in the Appendix \ref{TrottInt}. The final result is
obtained after summation over $ \, m $:
\begin{equation}\label{2ColIntFin}
   {\cal I}(k)= b \ { 1 \over \tilde{N}|\tau| b } \ \frac{\sqrt{ \beta \pi }}{2} \
                         { 1 \over  k! \, (k-1)! } \, .
\end{equation}

Substituting the combinatorial factor $ \, {\cal K}(k) $, Eqs.(\ref{GOEVcomb}-\ref{GUEVcomb}),
and the Trotter integral (\ref{2ColIntFin}) into the series (\ref{2ColSer}) we find:
\begin{eqnarray}\label{Ser2Col}
  \tilde{K}_1 & = &  2 \, \sqrt{\pi\beta} \
           \sum_{k=1}^{\infty} (-1)^k \, C^{(2)}_{\beta}(k) \, {\cal R}_N(k) \
                            ( \tilde{N} |\tau| b )^{2k-1} \, ; \\
  \label{Coef2ColGOE}
  C^{(2)}_{\beta=1}(k) & = &             { (2k-1)!! \over k! (k-1)! } \, ;   \\
  \label{Coef2ColGUE}
  C^{(2)}_{\beta=2}(k) & = &  { 1 \over (k-1)! } \, .
\end{eqnarray}
Eqs.(\ref{Ser2Col}-\ref{Coef2ColGUE}) correspond to the leading in $ \, b \ll 1 \, $
correction to the Poisson spectral form-factor. They have been obtained in the framework of the
two color approximation when we have taken into account only pairs of the interacting energy
levels. We note that the transformation (\ref{delta}) of the $ \, {\cal D}_N $--function
into the delta--function is approximate and its sub-leading terms yield the corrections to
$ \, b \tilde{K}_1 \, $ of higher orders starting from $ \, ~\sim b^3 ( \Delta \tilde{K}_1 ) $.

We emphasize the the formula (\ref{Ser2Col}) for $ \,\tilde{K}_1 \, $  is valid for a Gaussian
random matrix ensemble with a generic variance of the off-diagonal matrix elements
$ \, \langle |V_{ij}|^{2} \rangle =  b^{2}\,{\cal F}(|i-j|) $.

\section{The case of three colors}\label{3ColSec}

\begin{figure}[b]
\unitlength1cm
\begin{picture}(10,4.25)
   \epsfig{file=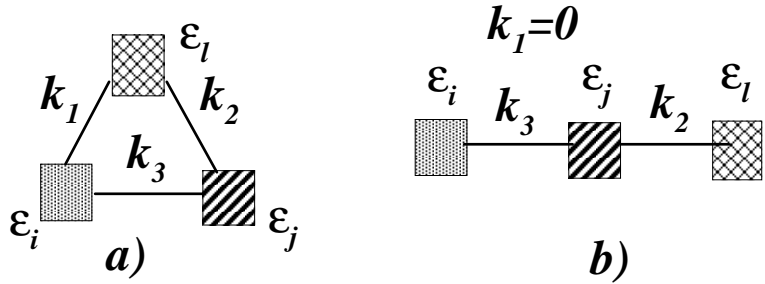,angle=0,width=10cm}
\end{picture}
\vspace{0.5cm}
\caption{
\label{3Col}
The three-color-diagrams: a) a triangle diagram with $ \, k_{1,2,3} > 0 $;
b) a line diagram with  $ \, k_{2,3} > 0 \, $ and $ \, k_{1} = 0 $.
}
\end{figure}

According to the selection rule explained above, the term of order $ \, b^2 \, $ in
$ \, \tilde{K}(\tau) \, $ results from an interaction of three energy levels, i.e., we have to
perform a summation of the three-color diagrams, see Fig.\ref{3Col}. Three shadowed
boxes stand for the three independent energy levels $ \, \varepsilon_{i}, \, \varepsilon_{j},
\, \mbox{ and } \varepsilon_{l} \, $ $ ( l > j > i ) $, which are connected by ILs.
Any two levels can be connected with each other independently of a connection with the
third one by an arbitrary number of ILs: $ \, k_1, \, k_2, \mbox{ and } k_3 $. This
leads to a three-dimensional power series in $ \, x = \tilde{N} | \tau | b $:
\begin{equation}\label{3ColSer}
   b^2 \tilde{K}_2 =
   \sum_{k_{1,2,3}}^{\infty} ({\bf i} \, x)^{2(k_1+k_2+k_3)} \ {\cal V} (\{k_{i}\}) \
             {\cal R}_N(\{k_{i}\}) \, ;  \quad i = 1, 2, 3 \, .
\end{equation}
Here, the universal factor $\, {\cal V} (\{k_{i}\}) \, $ contains the Trotter
integral $ \, {\cal I}(\{k_{i}\}) \, $ and the combinatorial factor $ \, {\cal K}
(\{k_{i}\}) $. The function $ \, {\cal R }_N \, $ is a generalization of the sum in real
space for the case of three colors:

\begin{equation}\label{RSS3Col}
   {\cal R}_N ( \{k_{i}\} ) = { 1 \over N } \, \sum_{l=j+1}^N \sum_{j=i+1}^N
       \sum_{i=1}^N
       \Biggl[
         \Bigl(  {\cal F}(|i-j|) \, \Bigr)^{k_1}
         \Bigl(  {\cal F}(|j-l|) \, \Bigr)^{k_2}
         \Bigl(  {\cal F}(|l-i|) \, \Bigr)^{k_3}
       \Biggr] \, .
\end{equation}

Generically, any energy level interacts with the two others and all
numbers of ILs are not zero $ \, k_{1,2,3} > 0 $. A diagram describing
this case will be referred to as {\it a triangle} (see Fig.\ref{3Col}A).
However, there are configurations where two levels do not interact with each
other and one parameter $ \, k_i \, $ is zero. A corresponding diagram will
be named {\it a line} (see Fig.\ref{3Col}B). We can attribute a certain
physical meaning to both of the three-color diagrams. Let us involve
a tree-like structure in the real space where a path with loops is prohibited.
It will generate only the ``lines''. On the contrary, a real space structure where
the paths with loops are  allowed will yield both the ``triangles'' and the ``lines''.

The combinatorial factor $ \, {\cal K } \, $ includes the number of  coloring of segments
by three colors (denoting $ \, \varepsilon_{i,j,l} $) and the number of ways
to pair $V$-matrices by in-circle and inter-circles connections. The $V$-matrices
create three sets of vertices in the circles: $ \, 2 k_1 \, $ vertices of $ \, \{
V_{i,l} , V_{l,i} \} $, $ \, 2 k_2 \, $ vertices of $ \, \{ V_{j,l} , V_{l,j} \} $,
and  $ \, 2 k_3 \, $ vertices of $ \, \{ V_{i,j} , V_{j,i} \} $. In view of
the property
\begin{equation} \label{IndPairing}
   \langle V_{s,p} V_{q,r} \rangle = \langle V_{s,p}^2 \rangle \, \delta_{s,q} \,
       \delta_{p,r} + \langle | V_{s,p} |^2 \rangle \, \delta_{s,r} \, \delta_{p,q}
\end{equation}
the vertices belonging to one and the same set must be paired with each other independently
of the two other sets. Applying the same arguments
that were used in the case of  two colors we  arrive at:
\begin{eqnarray}
   {\cal K}_{GOE} & = & Col_{GOE} \prod_{i=1,2,3} (2 k_i - 1)!! \, ; \\
   {\cal K}_{GUE} & = & Col_{GUE} \prod_{i=1,2,3} (k_i)! \, ;
\end{eqnarray}
where $ \, Col \, $ is the coloring factor. When the number of colors is larger than two
the coloring factor is different in GOE and GUE. Details of calculation of $ \, Col \, $
are given in Appendix \ref{AppCol}. Here we explain notations and give the results for GOE
and GUE.

Denote a total number of the segments for each of the three colors by $ \, R, \, G, \mbox{ and }
B $. These numbers can be expressed in terms of $ \, k_{1,2,3} $:
\[
  R = k1 + k2 \, ; \quad
  G = k2 + k3 \, ; \quad
  B = k3 + k1 \, .
\]
The colored segments are distributed over the two circles. The numbers of the colored segments
in the first circle will be labeled $ \, r, \, g, \mbox{ and } b $. Thus the second circle
contains $ \, R - r, \, G - g, \mbox{ and } B - b \, $ segments of different colors.

In the case of GOE, the two circles are colored independently under the only condition that
any two neighboring segments must be of the different colors (see Appendix \ref{AppCol}). The
factor $ \, Col_{GOE} \, $ reads
\begin{eqnarray}\label{Col3GOE}
 &   Col_{GOE} = Col^{(3)} ( r, g, b ) \, \times \, Col^{(3)} ( R-r, G-g, B-b ) \, ; \\
 &   Col^{(3)} ( a, b, c ) \equiv (a+b+c) \, 2^{a+b+c}
            \displaystyle
            \!\! \sum_{k=\max{(a,b,c)}}^{a+b+c \over 2} \!
            \frac{2^{-2k} \, (k-1)!}{ (k-a)! \, (k-b)! \, (k-c)! ( a+b+c-2k)! } \, .
    \nonumber
\end{eqnarray}

Before exploring the case of GUE, we remind that the ``crossed'' connections (see example 2 in
Fig.\ref{Vcon2C}) are not allowed in GUE. The ``crossed'' connections can be avoided if each set
of the vertices $ \, \{ V_{p,q}, \, V_{q,p} \} \, $ contains exactly $ \, k_i \, $ of $ \, V_{p,q}
\, $ vertices and $ \, k_i \, $ of $ \, V_{q,p} \, $ vertices regardless to their distribution
over the circles. Note, that this restriction applies only to the total number of $ \, V_{p,q}
\, $ and $ \, V_{q,p} \, $ while the number of the vertices $ \, V_{p,q} \, $ and $ \, V_{q,p}
\, $ in the same circle is not necessarily balanced. The coloring of the two circles is not
independent any longer: the difference $d=\, {\rm Number}[V_{p,q}] - {\rm Number}[V_{q,p}] \, $ of
the conjugated vertices must be the same in magnitude (but opposite in sign) for the two circles.
Therefore one has to compute the coloring factors for each ring at a fixed difference $ \, d \, $
and then to perform a  summation of the product of two coloring factors over $d$ (or,
equivalently, over the related variable $D=\frac{1}{2}(r+b+g-|d|)$). The result reads
(see Appendix \ref{AppCol} for the details):
\begin{eqnarray}\label{Col3GUE}
 &&   Col_{GUE} ( r, \, g, \, b, \, R, \, G, B ) \equiv \ l_1 \, l_2 \!\!
        \displaystyle \sum_{D=\max(r,g,b)}^{ l_1 \over 2} \!\! \Lambda(D) \times \cr
 &&    \Biggl\{  \displaystyle
          \sum_{p=\max(r,g,b)}^{D} { ( l_1-D-p, \, D-p, \, p-r, \, p-g, \, p-b) \over p } \times \\
 &&              \displaystyle
          \sum_{q=\max(R-r,G-g,B-b)}^{ \frac{l_2-l_1}{2} + D } \!\!
                 { \left( \frac{l_2+l_1}{2} - D -q, \, \frac{l_2-l_1}{2} + D - q, \,
                       q - (R-r), \, q - (G-g), \, q - (B-b)
                   \right) \over q }
      \Biggr\} \, ; \nonumber
\end{eqnarray}
\[
  l_1 \equiv r + g + b, \quad l_2 = R-r + G-g + B-b \, ;
\]
\[
   \Lambda(D) = \left\{
     \begin{array}{rcl}
        & 2 & \!\!, \ \mbox{ if } D < \displaystyle {l_1 \over 2} \, ; \\
        & 1 & \!\!, \ \mbox{ if } D = \displaystyle {l_1 \over 2} \mbox{ and } {l_1 \over 2}
                        \mbox{ is integer } \, .
     \end{array}
                \right.
\]
Brackets $ \, ( \alpha_1, \, \alpha_2, \, \alpha_2, \, \ldots ) \equiv \displaystyle \frac{ (
\alpha_1 + \alpha_2 + \alpha_3 + \ldots )! }{ \alpha_1! \, \alpha_2! \, \alpha_3! \, \ldots }
\, $ denote the multinomial coefficient.

The coloring factor has much simpler form for the line-diagram. For the example shown in
Fig.\ref{3Col}.b, where $ \, g = r + b; \, G = R + B $, $ \, Col \, $ acquires the following
form in GOE and GUE:
\begin{eqnarray}\label{LineCol}
  &  Col |_{ g = r + b; \, G = R + B }
             = Col_{line} ( r, \, b ) \ Col_{line} ( R-r, \, B-b ) \, ; \\
  &  Col_{line} ( a, \, b ) \equiv 2 \displaystyle { ( a + b )! \over a! \, b! } \, .
  \nonumber
\end{eqnarray}

There is no essential difference in the calculation of the Trotter integral for the
case of two and three
colors. We follow the algorithm explained in Appendix \ref{TrottInt}.
For {\it three}
colors, the integrand contains a product of {\it two} ``infusible'' $ \,  {\cal D}_N
\, $ functions.
They are converted to the $ \, \delta $-functions and the result of integration with
the accuracy of $ \, 1/(\tilde{N}\tau)^2 \, $ is
\begin{eqnarray}\label{Int3Col}
  & {\cal I} ( r, \, g, \, b, \, R, \, G, \, B )
                 = \displaystyle 2\pi \frac{\beta}{ \sqrt{3} } \ { 1 \over ( \tilde{N} \tau )^2 } \
           { 1 \over l_1 \, l_2 } \ \frac{ (R-2)! \, (G-2)! \, (B-2)! }{ ( R + G + B - 4 )! }
      \times \\
  &       \displaystyle \times { 1 \over ( r - 1 )! \, ( R - r - 1 )! } \
                             { 1 \over ( g - 1 )! \, ( G - g - 1 )! } \
                             { 1 \over ( b - 1 )! \, ( B - b - 1 )! }
            \, ; \nonumber
\end{eqnarray}
where the sum over all possible arrangements of the first Trotter segment is done.

To derive the universal factor $ \, {\cal V}( k_1, k_2, k_3) \, $ in Eq.(\ref{3ColSer}), we have to
perform summation over $ \, r, g, \mbox{ and } b $ accounting for the different distributions of
$V$-matrices over the traces. We remind that the matrices $ \, \hat V \, $ from the first trace
enter the Trotter chain with the minus sign. Therefore, the summand must be multiplied by
the factor $ \, (-1)^{l_1} $. Unlike the case of two colors, both the coloring factor
(\ref{Col3GOE},
\ref{Col3GUE}) and  the Trotter
integral (\ref{Int3Col}) depend on the distribution of the colored segments. Thus we arrive at the sum:
\[
   {\cal V}( k_1, k_2, k_3 ) =
   \sum_{r=1}^{R-1} \, \sum_{g=1}^{G-1} \, \sum_{b=1}^{B-1}
       \Bigl\{ (-1)^{ r + g + b } \
          {\cal K} ( r, \, g, \, b, \, R, \, G, \, B ) \times
          {\cal I} ( r, \, g, \, b, \, R, \, G, \, B ) \Bigr\} \, ,
\]
which turns out to be rather cumbersome. Amazingly, these summations can be done
analytically and we finally find the correction $ \, {\tilde K}_2 \, $ to the spectral
form-factor:
\begin{equation}\label{Ser3Col}
   {\tilde K}_2 = { \sqrt{3} \, \beta \over 3 } \!\! \sum_{k_1, \, k_2, \, k_3=0}^{\infty}
                      (-1)^{k_1+k_2+k_3} \ C^{(3)}_{\beta}(k_1, \, k_2, \, k_3) \,
                      {\cal R}_N (k_1, \, k_2, \, k_3) \ x^{ 2(k_1+k_2+k_3)-2} \, ;
\end{equation}
where ${\cal R}_N (k_1, \, k_2, \, k_3)$ is given by Eq.(33), $ \, x \equiv N | \tau | b \, $ and
the coefficients are
\begin{eqnarray}\label{Coef3ColGOE}
   C^{(3)}_{\beta=1} & = & - \, { \Gamma( k_1 + k_2 + k_3 )
                                  \over \Gamma( k_1 + k_2 + k_3 - 3/2 ) }
      \frac{ \Xi_{\beta=1}(k_1) \, \Xi_{\beta=1}(k_2) \, \Xi_{\beta=1}(k_3) }
           { \Gamma( k_1 + k_2 ) \, \Gamma( k_2 + k_3 ) \, \Gamma( k_1 + k_3 ) } \, ; \\
   \Xi_{\beta=1}( k ) & \equiv & \frac{ 2^k \, \Gamma(k-1/2) \, \Gamma(k+1/2) }
                                         { \sqrt{\pi} \Gamma(k +1) } \, ; \cr \cr
   \label{Coef3ColGUE}
   C^{(3)}_{\beta=2} & = & \left( 2 k_1 k_2 k_3 - k_1 k_2 - k_2 k_3
                               - k_1 k_3 \right) \
                           { \Xi_{\beta=2}(k_1) \, \Xi_{\beta=2}(k_2) \, \Xi_{\beta=2}(k_3)
                            \over \Gamma( k_1 + k_2 + k_3 - 3/2 ) } \, ; \\
   \Xi_{\beta=2}( k ) & \equiv & \frac{ \Gamma(k-1/2) }{ \Gamma(k+1) } \, .
   \nonumber
\end{eqnarray}
The correction $ \, {\tilde K}_2 \, $ is governed by the interaction of the three energy levels
and results from the summation of ``triangles'' and ``lines'':
All terms with $ \, k_{1,2,3} \ge 1 \, $ correspond to the ``triangle'' diagrams while the
``line'' diagrams are obtained by setting $ \, k_1 = 0 \, $ or $ \, k_2 = 0 \, $ or $
\, k_3 = 0 $.

The series in r.h.s. of Eq.(\ref{Ser3Col}) is three-dimensional and cannot be
reduced
to a product of one-dimensional series. Therefore an analysis of its behavior at arbitrary
$ \, x \, $ is not trivial. In the case of GUE, one can represent the function
$ \, \Gamma^{-1}( k_1 + k_2 + k_3 - 3/2 ) \, $ as an integral using the identity \cite{RizhGr}:
\begin{equation} \label{InvGamma}
   { 1 \over \Gamma( z ) } = \frac{1}{2\pi \, {\bf i}} \, \int_{-\infty}^{\infty} \,
                                     { \exp ( a + {\bf i} t)
                                \over ( a + {\bf i} t)^z } {\rm d} t \, , \ a > 0 \, ,
\end{equation}
and change the order of summations over $ \, k_i \, $ and integration over $ \, t $. The real-space
summation which is implied in the function $ \, {\cal R}_N \, $ has to be done at the last step.
This is an effective tool for the asymptotic analysis of the series (\ref{Ser3Col},\ref{Coef3ColGUE}).
The case of GOE is more complicated. Numerical and semi-numerical methods are useful here. We will
demonstrate an application of Eqs.(\ref{Ser3Col}-\ref{Coef3ColGUE}) to the Rosenzweig-Porter model
and to a critical almost diagonal PLBRM in the forthcoming paper \cite{next}.

\section{The case of crossover: almost unitary Gaussian ensemble}

We end the presentation of the method with the consideration of an almost unitary Gaussian RM
ensemble:
we explore the case of a crossover between GOE and GUE \cite{cross} which is close to GUE.
To define the almost unitary ensemble we introduce a parameter $ \,
\eta $ into Eq.(\ref{IndPairing}), which controls an asymmetry for the variance of
real and imaginary parts of the hopping
elements:
\begin{equation}\label{CrossDef}
   \langle V_{s,p} V_{q,r} \rangle = b^2 {\cal F} ( | s - p | ) \,
      \left[ \ \delta_{s,r} \delta_{p,q} \, + \, \eta \,
      \delta_{s,q} \delta_{p,r} \ \right] \, .
\end{equation}
In GOE, $ \, V_{s,p} \, $ is a real number and $ \, \eta_{GOE} = 1 $. In GUE the entries  $ \,
V_{s,p} \, $
are complex numbers, where $ \, \Re [ V_{s,p} ] \, $ and $ \, \Im [ V_{s,p} ] \, $ are
statistically independent random variables with equal variances and therefore $ \, \eta_{GUE}
= 0 $. Hence the parameter $ \, \beta \, $ is implicitly linked with $ \, \eta $. An ensemble of
RM is almost unitary if $ \, 0 < \eta \ll 1 $.

A first (trivial) impact of small $ \, \eta \, $ on the form-factor arises from the proportionality:
$ \, \tilde{K}_{1} \sim  \sqrt{\beta} \, ; \ \tilde{K}_{2} \sim \beta $ (see Eqs.(\ref{Ser2Col},
\ref{Ser3Col})), which comes from the Trotter integrals. We will concentrate on the other outcome
of a deviation from GUE where the non-zero parameter $ \, \eta \, $ leads to an appearance of new
series in powers of $ \, x $. We will derive a series that yields a leading in $ \, \eta $
correction to $ \, \tilde{K}_{1,2} |_{\beta=2} $.

When one performs the Gaussian averaging of $ \, V $-matrices, $ \, V$-vertices in the circles
can be paired by either ``crossed'' or ``non-crossed'' connections (see Fig.\ref{Vcon2C}). Each
``crossed'' connection is proportional to $ \, \eta \, $ in accordance to the definition
(\ref{CrossDef}). Therefore, the ``crossed'' connections are prohibited in GUE ($ \, \eta =
0 \, )$. To explore the leading term in $ \, \eta \ll 1 \, $ in the almost unitary case we have to
select the diagrams with a minimal number $ \, Cr \, $ of ``crossed'' ILs. Our main
task is to calculate their combinatorics.

\begin{figure}
\unitlength1cm
\begin{picture}(8,4.5)
   \epsfig{file=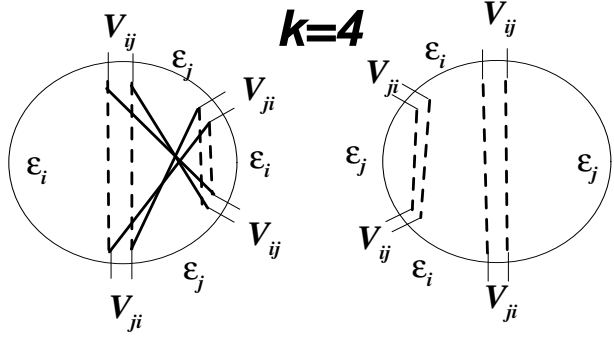,angle=0,width=8cm}
\end{picture}
\vspace{0.5cm}
\caption{
\label{VconCross}
Substitution of the ``non-crossed'' connections by the ``crossed'' ones: In GUE, only the
dotted ``non-crossed'' lines are allowed; the diagram with the minimal number of the
``crossed'' lines is obtained if one substitutes two dotted ILs by the solid ones.
}
\end{figure}

We start with the two colors. Let us consider a given distribution of $V$-matrices over the traces
at the fixed parameter $ \, k \, $ in GUE. There are sets of conjugated vertices: $
\, k \, $ vertices
$ \, V_{ij} \, $ and $ \, k \, $ vertices $ \, V_{ji} $; and there are $ \, k \, $ connections
without crossing. Each ``non-crossed'' connection pairs the vertex $ \, V_{ij} \, $ with
its conjugated counterpart $ \, V_{ji} \, $. A diagram with
the minimal number of ``crossed'' ILs is
obtained if we choose two ''non-crossed'' connections and rearrange them into two ``crossed''
connections, see an example in Fig.\ref{VconCross}. The rest $ \, k - 2 \, $ ILs remain
``non-crossed''. There are $ \, \Bigl( k(k-1)/2 \Bigr)^2 \, $ ways to choose two $ \, V_{i,j}$--
and two $ \, V_{j,i}$-vertices for the two ``crossed'' connections and $ \, (k-2)! \, $ ways to
draw the remaining ``non-crossed'' ILs. Therefore the combinatorial factor for the diagram with
the minimal number $ \, Cr = 2 \, $ reads:
\begin{equation}\label{CrossVcomb}
   {\cal K}_{\eta \ll 1} (k) = 4 \times { k \, (k-1) \over 4 } \ k! \, ,
\end{equation}
where the factor $ \, 4 \, $ accounts for two different ways to color each circle
by two colors. Substituting Eq.(\ref{CrossVcomb})
and the Trotter integral (\ref{2ColIntFin}) with $ \, \beta = 2 \, $ into the two-color series
(\ref{2ColSer}), we find the correction:
\begin{eqnarray}\label{Ser2ColCross}
  \Delta \tilde{K}_{1} & = & { \sqrt{2 \pi} \over 2 } \, \eta^2 \ \sum_{k=1}^{\infty} (-1)^k \,
                                                             { k \over (k-2)! } \, {\cal R}_N(k) \
                            x^{2k-1} \, ; \quad  x \equiv N | \tau | b  \, .
\end{eqnarray}

Each three-colored diagram in GUE has 3 sets of conjugated vertices, which numbers
are balanced: $ \, {\rm
Number} [V_{i,l}] = {\rm Number}[V_{l,i}]=k_1 $, $ \, {\rm Number}[V_{j,l}] = {\rm Number}[V_{l,j}]=k_2 $,
and $ \, {\rm Number}[V_{i,j}] = {\rm Number}[V_{j,i}]=k_3 $. To generate the diagram with the minimal
$ \, Cr = 2 \, $ we can repeat the arguments explained above for the case
of the two colors: a pair of ``non-crossed'' connections can be rearranged into a pair of
``crossed'' ones successively in the bonds $ \, V_{i,l} \leftrightarrow V_{l,i} \, ; \ V_{j,l}
\leftrightarrow V_{l,j} \, ; \mbox{ and } V_{i,j} \leftrightarrow V_{j,i} $. This procedure results in
the combinatorial factor
\begin{equation}\label{CrossVComb3}
   {\cal K} |_{\eta \ll 1 } = Col_{GUE}  \ { k_1(k_1-1) + k_2 (k_2 - 1) + k_3 (k_3 - 1) \over 4 } \,
                       \prod_{i=1,2,3} (k_i) \, ! \ ;
\end{equation}
with the coloring factor $ \, Col_{GUE} \, $ defined by Eq.(\ref{Col3GUE}).
Therefore the first correction to GUE takes the form:
\begin{equation}\label{Ser3ColCross}
   \Delta  {\tilde K}_2 = { 2 \, \sqrt{3} \over 3 } \, \eta^2 \!\! \sum_{k_1, \, k_2, \, k_3=0}^{\infty}
                      (-1)^{k_1+k_2+k_3} \ C^{(3)}_{\eta\ll1}(k_1, \, k_2, \, k_3) \,
                      {\cal R}_N (k_1, \, k_2, \, k_3) \ x^{ 2(k_1+k_2+k_3)-2} \, ;
\end{equation}
\begin{eqnarray}\label{Ser3ColCross2}
   C^{(3)}_{\eta\ll1} & = & { k_1 \, (k_1-1) + k_2 \, (k_2-1) + k_3 \, (k_3-1) \over 4 }
                            \times C^{(3)}_{\beta=2} \, .
\end{eqnarray}

Another way to insert the ``crossed'' ILs into the three-colored diagram in GUE is
to violate the
balance of the vertex numbers $ \, {\rm Number} [V_{i,l}] - {\rm Number}[V_{l,i}] = d_1 $,  $ \,
{\rm Number} [V_{j,l}] - {\rm Number}[V_{l,j}] = d_2 $, $ \, {\rm Number}[V_{i,j}] - {\rm Number}[V_{j,i}]
= d_3 $, $\,  d_{1,2,3} = \pm 1, \pm 2 \ldots $, and to compensate this violation by the ``crossed''
connections. One can prove that the integer parameters $ \, d_i \, $ are always equal to each other:
$ \, d_1 = d_2 = d_3  \equiv d \, $ (see Appendix \ref{AppCol}). Therefore the
minimal number of
the ``crossed'' connections obtained in this fashion is $ \, Cr |_{d = \pm 1} = 3 \, $ which is, however,
larger than $ \, Cr |_{d=0} = 2 \,  $ in Eqs.(\ref{Ser3ColCross}-\ref{Ser3ColCross2}).

\section{Examples}

In the last section, we give two examples of the application of the general
Eqs.(\ref{Ser2Col}-\ref{Coef2ColGUE}, \ref{Ser3Col}-\ref{Coef3ColGUE}, \ref{Ser2ColCross},
\ref{Ser3ColCross}-\ref{Ser3ColCross2}) for the corrections $ \, b \tilde{K}_1 \mbox{ and }
b^2 \tilde{K}_2 \, $ to the Poissonian form-factor $ \, \tilde{K}_P = 1 $. The examples
are the Rosenzweig--Porter model with the parameter $ \, b \, $ depending on the matrix size
$ \, N \, $ and the power law banded RM with the size-independent parameter $ \, b $.

\subsection{The Rosenzweig--Porter model}\label{exmplRP}

Let us consider the Rosenzweig--Porter model with
\begin{equation}\label{RPdef}
   b = { {\cal B } \over N } \, , \  {\cal B } \ll 1 \, ; \quad
   {\cal F}(i-j) = 1 \, .
\end{equation}
This definition corresponds to an almost diagonal RPRM where a weak perturbation of the diagonal
matrix can nevertheless yield a nontrivial level statistics \cite{ShapKunz}.

We note first, that the parameter $ \, x \, $ does not depend on the matrix size $ \, N$ :
\[
   x = \tilde{N} | \tau | b = \frac{\tilde{N}}{N} | \tau | {\cal B} \equiv T \, .
\]
The ratio $ \, \tilde{N}/N \, $ is a constant of the order of $ \, 1 $.
The product $ \, b^{c-1} {\cal R}_N ( \{k_i\} ) \, $ ($ \, c \, $ is the number of colors)
also has the finite limit at $ \, N \to \infty $. For example, in the case of {\it two colors}
\[
  \lim_{N \to \infty} \Bigl( b {\cal R}_N ( k ) \Bigr) =
  \lim_{N \to \infty} \Bigl( \frac{{\cal B}}{N} \ \frac{N-1}{2} \Bigr) = \frac{ {\cal B} }{2} \, ;
\]
and Eq.(\ref{Ser2Col}) takes the form:
\begin{equation}\label{Ser2ColRP}
  b \tilde{K}_1 |_{N\to\infty} =  \sqrt{\pi\beta} \ {\cal B} \
           \sum_{k=1}^{\infty} (-1)^k \, C^{(2)}_{\beta}(k) \ T^{2k-1} \, ,
\end{equation}
with the coefficients $ \, C^{(2)}_{\beta}(k) \, $ from Eqs.(\ref{Coef2ColGOE},\ref{Coef2ColGUE}).

Let us consider for simplicity the complex hopping elements, i.e., $ \, \beta = 2 $. The sum over
$ \, k \, $ in Eq.(\ref{Ser2ColRP}) can be easily calculated and we find:
\begin{equation}\label{RPansw1}
  b \tilde{K}_1(T, \beta = 2 ) |_{N\to\infty} =  - \sqrt{2\pi} \ {\cal B} \ T \, e^{-T^2} \, .
\end{equation}

In a similar manner we treat the case of {\it three colors}, where
\[
  \lim_{N \to \infty} \Bigl( b^2 {\cal R}_N ( k_1, k_2, k_3 ) \Bigr) = \frac{ {\cal B}^2 }{6} \, ;
\]
and consequently
\begin{equation}\label{Ser3ColRP}
   b^2{\tilde K}_2 |_{N\to\infty} = { \sqrt{3} \, \beta \over 18 } \ {\cal B}^2 \
                                    \sum_{s=2}^{\infty}
                                    (-1)^{s} \ {\rm C}_{\beta}(s) \ T^{2s-2} \, .
\end{equation}
The coefficient $ \, {\rm C}_{\beta}(s) \, $ is given by the sum:
\[
   {\rm C}_{\beta}(s) = \sum_{k1,k2,k3=0}^{\infty} C_{\beta}^{(3)} ( k_1, k_2, k_3 )
                           \delta_{ s - (k_1+k2+k_3) }
     = \sum_{s'=1}^{s} \sum_{k_3=0}^{s'} C_{\beta}^{(3)} ( s - s', s' - k_3, k_3 ) \, .
\]
Here we have introduced the sums of two and three indices: $ \, s
= k_1 + k_2 + k_3 \, , s' = k_2 + k_3 $; and $ \, C_{\beta}^{(3)}
\, $ is taken from Eqs.(\ref{Coef3ColGOE},\ref{Coef3ColGUE}). In
the case of the complex hopping elements, $ \, \beta = 2 $, all
summations over $ \, k_3, s', \mbox{ and } s $ in
Eq.(\ref{Ser3ColRP}) can be done analytically. The correction to
the form-factor takes a simple form:
\begin{equation}\label{RPansw2}
  b^2 \tilde{K}_2(T, \beta = 2 ) |_{N\to\infty} =
                        - \frac{2\sqrt{3} \, \pi}{9} \ {\cal B}^2 \ T^2 \,
                          ( 2 T^2 - 3 ) \, e^{-T^2} \, .
\end{equation}
The results (\ref{RPansw1}) and (\ref{RPansw2}) are in agreement
with the expansion in powers of $ \, {\cal B} \, $ of the exact
expression for the form factor $ \, \tilde{K}(T) \, $ obtained for
$ \, \beta = 2 $ in the paper [\onlinecite{ShapKunz}].

The case of the real hopping elements, $ \, \beta = 1 $, can also be analyzed based on 
the formulae (\ref{Ser2ColRP},\ref{Ser3ColRP}) but the calculations are lengthy and we 
present them in the forthcoming paper.

\subsection{Critical PLBRM}\label{exmplPLBRM}

 Before giving the example of a critical PLBRM, we would like to
discuss generic properties of the virial expansion in the case of
the $ \, N $-independent parameter $ \, b \, $ and a converging
real space sum Eq.(\ref{DandR}). In this case, all corrections to
the form factor $ \, \tilde{K}_m \, $ are given by the series in
powers of the parameter $ \, x = \tilde{N} |\tau| b $ that
diverges in the thermodynamic limit. In the leading in $ \, 1/N \,
$ approximation in Eq.(\ref{1/N}), where $ \, {\cal
R}_{N}(\{k_{i}\}) = {\cal R} (\{k_{i}\}) $, the entire dependence
of the spectral form-factor $\tilde{K}(\tau)$ on $\tau$ comes only
through the dependence on the parameter $x$. This means that
$\tilde{K}(\tau)$ is either $ \, \tau$-independent or divergent
(if the series in ${\tilde{N}|\tau| b}$ determines a function that
does not have a finite limit as $ \, {\tilde{N} |\tau|
b}\rightarrow\infty \, $) in the thermodynamic limit.

In order to obtain a dependence on $\tau$ for a {\it finite}
limiting spectral form factor $ \, \tilde{K}(\tau)|_{N\to\infty}
\, $ one has to account for $ \, 1/N $-corrections to the real
space sum $ \, {\cal R} (\{k_{i}\}) \, $ in Eq.(\ref{1/N}).
Rewriting $ \, 1/N \sim b |\tau| / (\tilde{N}|\tau| b) \, $ and
absorbing the factor $ \, (\tilde{N}|\tau| b) \, $ from the
denominator into the infinite series, we obtain the correction of
order $b\tau$ to the limiting spectral form factor:
\[
   \tilde{K}(\tau)|_{N\to\infty} \approx \chi_{0} + \, \chi_{1}\, b |\tau | + \ldots
\]
where $\, \chi_{0}(b)= 1+c_{01}\,b +c_{02}\,b^{2}+\ldots \, $ is
the unfolded level compressibility (see Eq.(\ref{K-chi})); $ \,
\chi_{1}(b)= c_{11}\,b + c_{12}\,b^{2}+\ldots $, and $ \, c_{ij}
\, $ are numerical coefficients. Thus the $ \, 1/N \, $ correction
to $ \, {\cal R} (\{k_{i}\}) \, $ determines $ \, \chi_{1} \, $
and leads to the $ \, b^{2}|\tau| \, $ correction in the limiting
form factor. For $ \, b\ll 1 \, $ this correction is small even at
the Heisenberg time $ \, \tau\sim 1 $. However, it could be
important for the {\it tail} of the two-level correlation function
Eq.(\ref{R-corr}) at large energy separations $ \, \omega \, $
because it is {\it non-analytic} in $ \, \tau $.

Now, we illustrate an application of the method developed using an
example of a critical almost diagonal PLBRM \cite{MF,ME}:
\begin{equation}\label{model2}
    b = {\cal B } \ll 1  \, ;          \quad
    {\cal F}( | i - j | ) \equiv { 1 \over 2 } \, \frac{1}{(i-j)^2} \, , \ i \ne j \, .
\end{equation}

We restrict ourselves to the case $ \, {\cal B} \tau \ll 1 \, $
omitting $ \, 1/N $-corrections to the real-space sum. The leading
term of $ \, {\cal R}_N (k) \, $ takes the following form:
\begin{equation}\label{RSS}
   {\cal R} (k) = \frac{1}{2^k} \, \lim_{N\to\infty}\sum_{m=1}^{N} \, \frac{1}{m^{2k}}  \, .
\end{equation}
Let us substitute expression (\ref{RSS}) into Eq.(\ref{Ser2Col})
and explore an asymptotic behavior $ \, x = \tilde{N} | \tau | {\cal B}
\to \infty \, $ of the series (\ref{Ser2Col},\ref{Coef2ColGOE})
and (\ref{Ser2Col},\ref{Coef2ColGUE}) for GOE and GUE
respectively. The simplest route is to change the order of the
summations:
\begin{equation}\label{Ser2Exmpl}
  {\tilde K}_{1}|_{{\cal B}\tau \ll 1} \simeq { 2 \sqrt{\beta \pi} \over N | \tau | } \
                    \sum_{m=1}^{\infty} \
                    \sum_{k=1}^{\infty} (-1)^k \, C^{(2)}_{\beta}(k) \
                    \left( {x \over m \sqrt{2}} \right)^{2k} .
\end{equation}
The summation over $ \, k \, $ yields:
\begin{equation}\label{Res2}
   {\tilde K}_{1} \simeq - { \sqrt{\beta\pi} \over N | \tau | }  \sum_{m=1}^{\infty}
     { x^2 \over m^2 } \exp\left( - { x^2 \over 2 m^2 } \right)
     \left\{
        \begin{array}{l}
          I_0 \left( { x^2 \over 2 m^2 } \right) - I_1 \left( { x^2 \over 2 m^2 } \right) \, , \
                   \beta = 1 \, ; \cr
          1 \, , \ \beta = 2 \, .
        \end{array}
     \right.
\end{equation}
Here $ \, I_{0,1} ( \cdots) \, $ are the Bessel functions.  The
sum over $ \, m \, $ converges at $ \, m \sim x \gg 1 \, $
therefore it can be converted to the integral over $ \, m $. After
this integration we find the two-colors correction to the level
compressibility $ \, \chi_0 \simeq 1 + {\cal B}\times c_{01} $:
\begin{equation}\label{Result2}
   c_{01} |_{\beta = 1} = -2 \, ; \quad c_{01} |_{\beta = 2} = -\pi \, .
\end{equation}
Note, that Eq.(\ref{Result2}) can be reproduced with the help of a
heuristic renormalization--group method (compare with
[\onlinecite{ME}]).

Using our method it is also easy to derive from
Eq.(\ref{Ser2ColCross}) the correction in the almost unitary case:
\[
   \Delta c_{01} |_{\eta \ll 1} = \eta^2 { \pi \over 16 } \, .
\]

We emphasize that it is the behavior of $ \, {\cal F}(|i-j|) \, $
at large $ \, |i-j| \, $ that affects the factor $ \, {\cal R}(k)
\, $ and governs the asymptotic behavior of the corresponding
series. Depending on it, the two-color correction may either
approach a constant limit or diverge as $ N \rightarrow\infty$.
One can show \cite{next} that: 1) for ${\cal F}(|i-j|)$ decreasing
{\it faster} than $1/|i-j|^{2}$ the two-color correction {\it
vanishes} in the limit $ \, N \rightarrow\infty $; 2) for  ${\cal
F}(|i-j|)$ decreasing {\it slower} than $1/|i-j|^{2}$ it is {\it
divergent}. This is the manifestation of the
localization-delocalization transition in the spectral statistics.
Note that even the case, where the series in r.h.s. of
Eq.(\ref{Ser2Col}) does not have a finite limit as $ \, N \to
\infty $, can be considered by our formalism as long as the
correction $ \, \tilde{K}_1(\tau) \, $ is small at a finite $ \, N $.

A  remarkable feature of the critical PLBRM with $ \, {\cal F}
\sim 1/|i-j|^{2} \, $ in the {\it three-color approximation} is the 
logarithmic divergence $ \, {\cal B}^2 \log^2 ( N | \tau | {\cal B} ) \, $ 
of the line- and the triangle-diagrams (see Fig.\ref{3Col}) at any $ \,
\beta $. However, in the sum of the line- and the
triangle-diagrams, the logarithmically divergent terms of order
$ \, {\cal B}^{2} \, $ cancel out for both GOE and GUE  so that $ \,
\lim_{x\to\infty} \Bigl( \tilde{K}_2(\beta=1,2) \Bigr) \, $ is a
finite constant of order 1. The sub-leading diverging term  {\it 
survives strikingly in the three colors correction of the almost unitary 
ensemble}: $ \, {\cal B}^2 \Delta K_2 \sim \, - {\cal B}^{2} 
\eta^2 \log ( N | \tau | {\cal B} ) $. This  
indicates a failure of the virial expansion for the critical PLBRM as
$N\rightarrow\infty$ in the crossover between the unitary and
orthogonal ensembles and raises a question on the cancellation of
the logarithmically divergent corrections of higher orders in $b$
in a pure GOE or GUE. This question deserves a separate detailed
study elsewhere.

\section{Conclusions}

In this paper, we have developed a new method to study the spectral statistics of Hermitian
Gaussian random matrices with parametrically small hopping elements $ \, H_{i,j} \,
$
as compared to the diagonal ones: $ \, \langle | H_{i \ne j} |^2 \rangle / \langle
\varepsilon_k^2
\rangle \sim b^2 \ll 1 $. We have derived a regular virial expansion of the spectral form-factor
in powers of $b$, where the virial coefficient in front of $b^{{\cal A}}$ is given by interaction
of ${\cal A}+1$ levels.

The expansion is represented by diagrams which are generated with the help of the Trotter
formula. We have established the rigorous selection
rule for the diagrams, which allows to account for exact contributions of a given number of resonant
and non-resonant interacting levels. Thus the method offers a controllable way to find an answer to the
question when a weak interaction of levels can  drive the system from localization
toward criticality and delocalization.

The method applies to the spectral properties of the random matrices with uncorrelated entries and
with a generic dependence of the variance $ \, \langle | H_{i,j} |^2 \rangle \equiv b^2 {\cal F}( | i -
j | ) \, $ on the distance $ \, | i - j | \, $ from the main diagonal. We calculated the corrections
governed by  interaction of two, Eqs.(\ref{Ser2Col}-\ref{Coef2ColGUE}), and three,
Eqs.(\ref{Ser3Col}-\ref{Coef3ColGUE}), levels in the cases of GOE, GUE and in the almost unitary ensemble,
Eqs.(\ref{Ser2ColCross},\ref{Ser3ColCross}-\ref{Ser3ColCross2}), for an arbitrary function $ \, {\cal F}
( | i - j | ) $. These equations are the main results of the paper. We have demonstrated their application
to an almost diagonal Rosenzweig--Porter model and to an almost diagonal power law banded random
matrices.

The calculation of the virial coefficients is based on the solution to the combinatorial
problem of the graphs coloring. The solution for the coloring of a single graph
is known in a closed form for an arbitrary number of colors \cite{BK}. Thus using our method
one can, in principle, compute the virial coefficient of an arbitrary order.

\begin{acknowledgments}

We are very grateful to Igor Krasovsky, Boris Shapiro and Riccardo Zecchina for useful discussions, to
Holger Schanz for his help in combinatorial analysis, and to Denis Basko for careful reading this paper.

\end{acknowledgments}

\appendix

\section{The level compressibility}\label{AppDef}

In this Appendix, we derive an approximate relation between the level compressibility and
the correlation function $ \, R $. Let us define the two-level correlation function
$ \, R_{E}(\omega) \, $ centered at an energy $ \, E $:
\begin{equation}\label{R-redef}
   R_E (\omega) = \frac{\langle\langle \, \sum_{n,m} \delta (E-\epsilon_n) \, \delta (E-\omega-\epsilon_m)
                         \, \rangle\rangle}{\langle \rho(E) \rangle^2}
\end{equation}
In this definition it is assumed that $\omega$ is small compared to the total
bandwidth, so that $\langle \rho(E)\rangle\approx \langle \rho(E-\omega)\rangle$.

We set apart the diagonal (singular) part in the double
sum in numerator of (\ref{R-redef}):
\begin{equation} \label{R-sep}
  R_E (\omega) = R^{(r)}_E (\omega) + \frac{\delta(\omega)}{\langle \rho(E) \rangle} \, .
\end{equation}
Then the normalization by $\langle \rho(E) \rangle^2$ ensures that the {\it regular}
part $ R_{E}^{(r)}
(\omega)$ depends very weakly on the energy $E$, except for regions of the
width $\sim \omega$ near the band edges.

The level number variance can be expressed in terms of
the correlation function $ \, R (\omega) $~[\onlinecite{CKL}]:
\begin{equation}\label{VarE}
  \Sigma_2 (\bar n) = \int_0^{\bar n \Delta} \int_0^{\bar n \Delta}
        \left( \langle \rho(E) \rangle^2 R_E (E-E') \right) \, {\rm d} E \, {\rm d} E' \, .
\end{equation}

Substituting decomposition (\ref{R-sep}) into Eq.(\ref{VarE}),
one finds
\begin{equation}\label{VarEAppr}
  \Sigma_2 (\bar n) \simeq \bar n + \int_{- \bar n \Delta}^{\bar n \Delta} { \bar n \Delta - | \omega |
       \over \Delta } \, R^{(r)} (\omega) \, { {\rm d} \omega \over \Delta } \, ; \quad
    R^{(r)} (\omega) \equiv R^{(r)}_{E=0} (\omega) \, .
\end{equation}
This expression is valid up to the leading order $ \, O(b) \, $ with respect to $ \, b \ll 1 $.

The definition (\ref{chi-def}) for the level compressibility yields now the following expression:
\begin{equation}\label{chi-int}
  \chi = 1 + \lim_{\bar n \to \infty} \int_{- \bar n }^{\bar n } {\rm d}s
          \, \lim_{N \to \infty} R^{(r)} (s \Delta) \, .
\end{equation}

Using definition (A1), formula (\ref{R-sep}) and the identity
\[
  \int_{-\infty}^{+\infty}\, {\rm d}E \int_{-\infty}^{+\infty}\, {\rm d}\omega \,
      \langle\langle \rho(E) \rho(E-\omega) \rangle\rangle \, e^{{\bf i}\omega t} \equiv
  \langle\langle \, {\rm Tr} \, e^{{\bf i} \hat H t } \ {\rm Tr} \, e^{-{\bf i} \hat H t } \,
  \rangle\rangle
\]
we derive a useful relation:
\begin{equation}\label{Tr-R}
  {1 \over N} \langle\langle \, {\rm Tr} \, e^{{\bf i} \hat H t } \ {\rm Tr} \, e^{-{\bf i} \hat H t } \,
              \rangle\rangle \simeq
  1 + \Upsilon \int_{- \infty}^{+ \infty} {\rm d}s  \,
e^{{\bf i}s t\Delta} R^{(r)}
  (s\Delta) \, .
\end{equation}
which remains valid also in the limit $N\rightarrow\infty$.

At $ \, t = t_0 \equiv 1 / ( \bar n \Delta ) $, large energy scale
$ \, \omega \equiv s\Delta \gg \bar n \Delta \, $
does not essentially contribute to the integral in Eq.({\ref{Tr-R}) due to the strong
oscillations of the
integrand. Hence, we have:
\[
   \lim_{\bar n \to \infty} \int_{- \bar n }^{\bar n } {\rm d}s
          \, \lim_{N\rightarrow\infty} R^{(r)} (s\Delta) = \lim_{\tau_0 \to 0}
\int_{- \infty}^{+ \infty}
               {\rm d}s \, e^{{\bf i}s \tau_0 } \lim_{N\rightarrow\infty} R^{(r)}
  (s\Delta) \, ; \quad \tau_0 \equiv t_0 \Delta \, ,
\]
and using Eqs.(A5,A6) arrive at the expression (\ref{K-chi}).

\section{Integration over the Trotter variables}\label{TrottInt}

In general, the Trotter integral depends  on the number of colors and on
the distribution of
matrices $ \, \hat V \, $ over the traces. The former defines the number of
`infusible' functions
$ \, {\cal D}_N \, $ in the integrand while the number of the Trotter variables is
determined by the latter. In this Appendix, we calculate the integral in the case of 2 colors at
an arbitrary number of the colored segments in the circles determined by the parameter $ \, m $.
The integrand in Eq.(\ref{TrI-2}) (see the Section \ref{IntRig}) contains two additive parts
reflecting two different possibilities to color the first segments. Unlike that example,
we will use another procedure with a simplification that we will not pay an attention to
a specific color of the first segments. The final result of the calculation is not affected
by this simplification.

Let us apply the Trotter formula to both traces in r.h.s. of Eq.(\ref{K0}) involving two
infinitely long Trotter chains of the length $ \, n_{1,2} \to \infty $ and let the number
of the colored segments in the circles be $ \, 2 m \, $ and $ \, 2 ( k - m ) $. We remind:
correct coloring by 2 colors imposes a restriction that a circle must contain equal number
of sectors for each color. Therefore, the total number of the segments for a given color
is $ \, k \, $ with $ \, m \, $ segments belonging to the first circle and $ \, k - m \, $
segments in the second one.

Denote the discrete Trotter variables coming from the first trace as $ \, y_1, \, y_2, \,
\ldots , \, y_{2m} \, $ so that the variables $ \, y_1, \, y_2, \, \ldots , \, y_{m} \, $
mark the segments of the first color while the rest of variables are attributed to the
segments of the second color. The variables from the second trace are denoted as $ \, z_1, \,
z_2, \, \ldots z_{k-m} \, $ for the first color and $ \, z_{k-m+1}, \, z_{k-m+2}, \,
\ldots z_{2(k-m)} \, $ for the second one. Since two neighboring subsegments separated by the
origin of the Trotter chain (instead of the $V$-vertex) are of the same color we fuse them and
mark by the single Trotter variable.

We rescale the variable introducing the continuous ones: $ \, Y_i = y_i /
n_1 \, , \  Z_i = z_i / n_2 $. They are normalized as (see Eq.(\ref{norm}))
\begin{equation}\label{NormGen}
   \sum_{i=1}^{2m} Y_i = \sum_{i=1}^{2(k-m)} Z_i = 1 \, ; \quad
      Y_i \ge 0 \, , \ Z_i \ge 0 \, .
\end{equation}
The integrand must be multiplied by the factor:
\begin{equation}\label{Pref}
   \frac{ n_1 }{ 2 m } \ \frac{ n_2 }{ 2 ( k - m ) } \, .
\end{equation}
The numerator reflects an invariance with respect to a cyclic permutation of the
matrices under the traces while the denominator excludes multiple counting of
the equal configurations of $V$-vertices in the circle.

We follow further the algorithm explained in the Section \ref{IntRig} and, accounting
for the property (\ref{NormGen}) and fusing two $ \, {\cal D}_N \, $ functions as it is
done in Eq.(\ref{TrI-2}), the Trotter integral can be presented as:
\begin{eqnarray}\label{Cmk}
   {\cal I}(k,m) & = & \frac{ 1 }{ 2 m } \ \frac{ 1 }{ 2 ( k - m ) } \
       \int_0^1 \prod_{i=1}^{2m-1} {\rm d} Y_i \
       \int_0^1 \prod_{j=1}^{2(k-m)-1} \!\!\! {\rm d} Z_j \ \times \cr
   \label{Imk}
       & \times &
      \ \theta \left( 1 -  \sum_{i=1}^{2m-1} Y_i \right) \
            \theta \left( 1 -
                \!\!\! \sum_{j=1}^{2(k-m)-1} \!\!\! Z_j \right) \
         {\cal D}_N \left( \sqrt{2} \sum_{i=1}^m Y_i -
           \sqrt{2} \sum_{j=1}^{k-m} Z_j \right) \, .
\end{eqnarray}
Having integrated over all variables but $ \, \widetilde{Y} = \sum_{i=1}^{m} Y_i \, ; \
\widetilde{Z} = \sum_{j=1}^{k-m} Z_j $, we arrive at
\begin{eqnarray}\label{IntSimp}
      {\cal I}(k,m) & = & \frac{ 1 }{ 4 }
   \int_0^1
   { \widetilde{Y}^{m-1} (1-\widetilde{Y})^{m-1} \over m!(m-1)! }
   {\rm d} \widetilde{Y} \!\!\!
   \int_0^1
   { \widetilde{Z}^{k-m-1} (1-\widetilde{Z})^{k-m-1} \over (k-m)!(k-m-1)! }
   {\rm d} \widetilde{Z} \
         {\cal D}_N ( \sqrt{2} [ \widetilde{Y} - \widetilde{Z} ] ) \\
                    & \simeq &
   \sqrt{ \beta \pi } \ { 1 \over \tilde{N} |\tau| } \
   \frac{ 1 }{ m!(m-1)! (k-m)!(k-m-1)! } \
   { (k-2)! \over 2^{k} \times (2k-3)!! } \, .
                 \label{IntFin}
\end{eqnarray}
At the last step, we sum over the parameter $ \, m \, $ to account for different
distributions of $V$-matrices over the traces:
\begin{equation} \label{2ColSum}
   {\cal I} (k) = \sum_{m=1}^{k-1} {\cal I}(k,m) = { \sqrt{ \beta \pi } \over 2 } \
       { 1 \over \tilde{N} |\tau| } \ { 1 \over k! \, (k-1)! } \, .
\end{equation}

\section{Combinatorial coefficient of coloring}\label{AppCol}

To calculate the universal factor $ \, {\cal V} ( \{ k_i \} ) \, $ at any number of colors
$ \, c \, $ we need a combinatorial factor of the circles coloring: $V$-vertices divide the
circles into the segments. The colors, which denote the statistically independent energy levels
$ \, \varepsilon_i $, should be distributed over the segments by all possible ways under the
condition that all adjacent segments must bear different colors.

Let us start with the coloring of a single circle. The question of our interest was formulated in
Appendix of paper [\onlinecite{BK}] as follows: How many permutations of $ \,  L \, $ objects of
$ \, c \, $ different colors (given $ \, g_1 \, $ objects of the first color, $ \, g_2 \, $ objects
of the second color, etc.) are there under the condition that no objects of the same color may stand
next to each other? We have to supplement it with two additions: 1) we do not distinguish the segments
of the same color; 2) we can do cyclic permutations of the segments in the circle. Thereafter, the
generic coloring factor can be obtained easily from Eq.(30) of paper [\onlinecite{BK}]:
\begin{eqnarray}\label{Berk}
   P_{ g_1, \, g_2, \, \ldots , \, g_c } (c) & = & L \, (-1)^{L-c} \sum_{n=0}^{\infty} (-1)^n
         \frac{\partial^{c-1+L}}{\partial \, x^{c-1+L}}
         \left[  {1 \over x} \, \prod_{i=1}^c h_{g_i}(x)
         \right] \Biggl|_{x=0} \, ; \\
                \label{BerkAux}
   h_g(x) & = & \sum_{s=1}^g \, \left(
                     \begin{array}{c}
                         g-1 \cr
                         g-s
                     \end{array}
                                \right)
                \, { x^s \over s! } \, .
\end{eqnarray}
Eqs.(\ref{Berk}-\ref{BerkAux}) describe any situation with the arbitrary number of colors and the arbitrary
number of the segments. Unfortunately, these expressions are cumbersome and are not convenient for a further
calculation of the factor $ \, {\cal V} $. In this Appendix, we will derive a simpler formula for the
particular case of three colors, $ \, c = 3 $.

\begin{figure}
\unitlength1cm
\begin{picture}(8.0,3.5)
   \epsfig{file=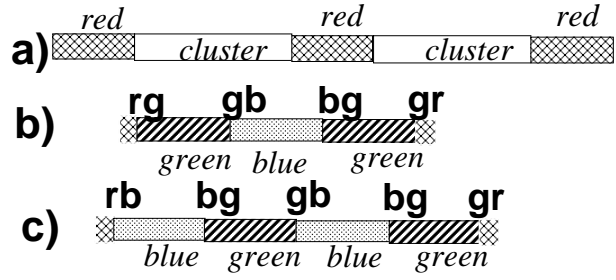,angle=0,width=8.0cm}
\end{picture}
\vspace{0.5cm}
\caption{
\label{Clusters}
a) An unrolled circle with $ \, r = 2 \, $ and with two clusters of green and blue segments;
b) A cluster of odd length; c) A clusters of even length. Pairs of bold letters denote
switching of colors at the $V$-vertices.
}
\end{figure}

We use the notations from Section\ref{3ColSec} labeling the numbers of the colored segments by $ \, r =
g_1, \, g = g_2, \, \mbox{ and } b = g_3 $. The first, the second and the third
colors will be denoted as
``red'', ``green'' and ``blue''. Let a color of the first segment be, for example, red.
There are $ \, r \, $ red segments which separate out $ \, r \, $ clusters. The clusters contain
green and blue segments, which must be arranged in an alternating fashion. A number of the segments in
the cluster will be referred as a ``length of the cluster''. The are different types of the clusters
(see Fig.\ref{Clusters}): 1) The clusters with an even length that enclose pairs of sectors; 2)~The
clusters with an odd length that enclose the pairs as well as one extra segment (either green or blue).
We denote a number of these clusters accordingly: $ \, n_0, \, n_g, \mbox{ and } n_b $. The parameters
$ \, n_{0,g,b} \, $ are connected to the numbers of the segments:
\begin{equation}
   n_g + n_0 + n_b = r \, ; \quad
   g - b = n_g - n_b   \, .
\end{equation}

First we have to distribute the clusters of three types over $ \, r \, $ positions. The number of
the distributions is given by the multinomial coefficient:
\begin{equation}\label{Comb1}
   ( n_0, \, n_g, \, n_b ) \equiv \frac{r!}{n_0! \, n_g! \, n_b!} \, .
\end{equation}
Given that each of $ \, n_0 \, $ clusters of even length contains at least one pair
we distribute $ \, { b - n_b - n_0 = g - n_g - n_0 } \, $ pairs of the segments over $ \, r \, $
clusters. The corresponding combinatorial factor is:
\begin{equation}\label{Comb2}
  2^{n_0} \,
  \left(
     \begin{array}{c}
       r + \Bigl( b - n_b - n_0 \Bigr) - 1 \cr
       r - 1
     \end{array}
  \right) \, ,
\end{equation}
where $ \, 2^{n_0} \, $ reflects the possibility to start the clusters of even length with either
a green segment or with a blue one.

The coloring factor $ \, Col_r ( r, g, b ) \, $ for one circle with the red first
segment is obtained after summation of the product of Eq.(\ref{Comb1}) and Eq.(\ref{Comb2})
over the independent number of the clusters, for example, over the parameter $ \, n_b \, $
in the range $ \, 0 \le n_b \le \left[ { r + b - g } \over 2 \right] $.
To find the second $ \, Col_g ( r, g, b ) \, $ and the third $ \, Col_b ( r, g, b ) \, $ parts, we
have to repeat the same procedure assigning  green and blue colors to the first
segment of the circle. Eq.(\ref{Col3GOE}) has been obtained after the summation of all parts\cite{Sch}:
\[
  Col^{(3)} ( r, g, b ) = Col_r ( r, g, b ) + Col_g ( r, g, b ) + Col_b ( r, g, b ) \, .
\]

Eq.(\ref{Col3GUE}), that is used to calculate the universal factor $ \, {\cal V} ( k_1, k_2, k_3 ) \, $
at $ \, c = 3 \, $ in GUE, has been obtained under an additional condition. Namely, to exclude the
``crossed'' connections of the $V$-vertices each set $ \, \{ V_{p,q}, \, V_{q,p} \} \, $ must contain an
equal number of the ``conjugated'' vertices: $ \, {\rm Number}[V_{p,q}] = {\rm Number}[V_{q,p}] \, $ (see
Section \ref{3ColSec}). This restriction applies only to the total number of $ \, V_{p,q} \, $ and
$ \, V_{q,p} \, $ while the number of the conjugated vertices in the same circle is not necessarily
balanced. Therefore, coloring of the two circles is not independent.

Let us consider the balance between $ \, V_{p,q} \, $ and $ \, V_{q,p} \, $ in more detail. The
conjugated vertices inside a cluster with an odd length and on its boundary  are always balanced (see
an example in Fig.\ref{Clusters}b). Hence, the clusters of odd length cannot violate the total
balance. It is not true for a cluster of an even length: the imbalance between the conjugated
vertices is $ \, \pm 1 \, $ for all three sets: $ \, \{ V_{i,l},  V_{l,i} \}; \, \{ V_{j,l},  V_{l,j} \};
\mbox{ and } \{ V_{i,j},  V_{j,i} \} $ (see an example in Fig.\ref{Clusters}c). The sigh of the imbalance
depends on a color of the first segment of the given cluster.

Denote the number of clusters of an even length, which start with
green and blue segments, by $ \, n_0^{(-)} \, $ and $ \, n_0^{(+)}
= n_0 - n_0^{(-)} \, $ respectively and introduce a difference $
\, d = n_0^{(+)} - n_0^{(-)} = n_0 - 2 n_0^{(-)} \, $ which will
be referred to as ``the number of defects''. It gives the number
of imbalanced conjugated vertices of each set in a given circle.
Below, the upper indices in $ \, d^{(1)} \, $ and $ \, d^{(2)} \,
$ will label the number of defects in the first and in the second
circles correspondingly.

Since each cluster of even length produces the same imbalance between $ \, V_{p,q} \, $ and $ \, V_{q,p}
\, $ in all sets of the vertices, the number of the conjugated vertices is balanced if
\begin{equation}\label{ballance}
  d^{(1)} + d^{(2)} = 0 \, .
\end{equation}
Equality (\ref{ballance}) must hold true at any combination of the colors of the first segments and
expresses the difference of the coloring in GOE and GUE. Eq.(\ref{Col3GOE}), which is used in GOE, results
from an independent summation over the number of defects $ \, d_1 \, $ and $ \, d_2 \, $ in each circle.
In GUE, we calculate the coloring factor $ \, \widetilde{Col}_r \, $ of one circle at the fixed
$ \, d \, $ and perform the summation over the number of defects obeying Eq.(\ref{ballance}):
\[
   \sum_{d^{(1)}} \sum_{d^{(2)}} \,
                 \Bigl\{ \,
                         \widetilde{Col}_r ( r, g, b, d^{(1)} ) \times
                         \widetilde{Col}_r ( R - r, G - g, B - b, d^{(2)} ) \times
                         \delta_{d^{(1)},-d^{(2})} \,
                 \Bigr\} \, .
\]
The summand is the product of the coloring factors for the two circles.

To find the expression for $ \, \widetilde{Col}_r \, $ we have to sum the product of Eqs. (\ref{Comb1})
and the factor similar to Eq.(\ref{Comb2}):
\[
    \left(
     \begin{array}{c}
       n_0 \cr
       n^{(-)}
     \end{array}
  \right) \times
  \left(
     \begin{array}{c}
       r + [ b - n_b ] - n_0 - 1 \cr
       r - 1
     \end{array}
  \right) \, ;
\]
over the parameter $ \, n_b $. The factor $ \, \left( \begin{array}{c} n_0 \cr n^{(-)} \end{array}
\right) \, $ with $ \, n^{(-)} = \displaystyle \frac{ n_0 - d }{2} = 0, 1, \ldots \, $ accounts for
the different distributions of the clusters of the even length starting with green and blue colors.
Eq.(\ref{Col3GUE}) results from the summation over all combinations of the colors of the first segments.

\end{document}